\documentclass[
twocolumn,
amsmath,amssymb,
aps,
physrev,
prstab,
]{revtex4-2}

\usepackage[colorlinks=true, linkcolor=blue, citecolor=blue, urlcolor=blue]{hyperref}
\usepackage{graphicx}
\usepackage{dcolumn}
\usepackage{bm}
\usepackage{booktabs}

\begin{document}

\preprint{APS/123-QED}

\title{\textbf{Physics design of a dual-ring light source   toward MHz-rate free-electron laser radiation} 
}%

\author{Fancong Zeng}
\affiliation{  Institute of High Energy Physics,
Chinese Academy of Sciences, and University of Chinese Academy of Sciences, Beijing 100049, China}

\author{Yi Jiao}
\email{  jiaoyi@ihep.ac.cn}
\affiliation{  Institute of High Energy Physics,
Chinese Academy of Sciences, and University of Chinese Academy of Sciences, Beijing 100049, China}

\author{Weihang Liu}
\email{  liuwh@ihep.ac.cn}
\affiliation{ Institute of High Energy Physics,
Chinese Academy of Sciences, and University of Chinese Academy of Sciences, Beijing 100049, China}
\affiliation{Spallation Neutron Source Science Center, Dongguan 523803, China} 

\author{Yu Zhao  }
\affiliation{  Institute of High Energy Physics,
Chinese Academy of Sciences, and University of Chinese Academy of Sciences, Beijing 100049, China}
\affiliation{Spallation Neutron Source Science Center, Dongguan 523803, China}

\author{Weilun Qin }
\affiliation{  Institute of High Energy Physics,
Chinese Academy of Sciences, and University of Chinese Academy of Sciences, Beijing 100049, China}
\affiliation{Spallation Neutron Source Science Center, Dongguan 523803, China} 
 
\begin{abstract}

Combining the high peak brilliance  and short pulse duration of free-electron lasers
(FELs) with the high repetition rate of diffraction-limited storage rings (DLSRs) is a promising direction for accelerator light sources. In this paper, we propose
a dual-ring light source concept that, at MHz repetition rates, delivers pulses with a
pulse durations and peak power intermediate between those of FELs and DLSRs. 
Such a light source can be realized as an upgrade of an existing DLSR by adding an isochronous ring equipped with transverse gradient undulators (TGUs) and connecting  the two rings via bunch compression and decompression systems. 
  We present the conceptual layout and show that
bunch compression shortens the electron bunches to ps-scale durations; multi-turn
radiation in the isochronous ring  raises the
repetition rate to 1.36~MHz  from a single TGU; and the long TGUs enhance the
brilliance, delivering $10^{7}$--$10^{9}$ photons per pulse across the soft and hard
X-ray range, an increase of 2--4 orders of magnitude over the original DLSR.
The dual-ring light source thus provides a cost-effective route to MHz FEL operation based on existing
storage-ring infrastructure, combining the high repetition rate of a DLSR with the short
pulses and high brilliance of a linac-driven FEL.

\end{abstract}


\maketitle

\section{Introduction}


Accelerator-based light sources, including  ring-based synchrotron radiation facilities and linear accelerator (linac)-based free-electron lasers (FELs),  are indispensable tools that have enabled groundbreaking discoveries across diverse research fields.
Driven by growing user demands, storage-ring-based light sources have evolved from third- to fourth-generation, i.e., diffraction-limited storage rings (DLSRs), as exemplified by MAX-IV, APS-U, ESRF-EBS, HALF, and HEPS~\cite{Tavares:2018,Sajaev:2019,Jiao:2018,Raimondi:2023,Liu:2014,Bai:2026}. These facilities offer exceptional stability, tens of beamlines, and repetition rates of $\sim$100 MHz. However, their long bunches limit the temporal resolution to $20$--$200$ ps, and the per-pulse photon yield is typically $10^{4}$--$10^{6}$.
Linac-driven FELs, by contrast, operate in a single-pass configuration, 
 allowing electron bunches to be compressed to  attosecond-to-femtosecond durations with kiloampere-level peak currents that deliver $10^{11}$--$10^{13}$ photons per pulse. Nevertheless, their repetition rate is constrained to below 120 Hz with normal-conducting linac technology \cite{Emma:2010,Ishikawa:2012,Hara:2013,Allaria:2012,Kang:2017,Prat:2020}. Superconducting linac-driven FELs, such as LCLS-II(HE) and SHINE, can reach MHz repetition rates, but at substantially higher cost and with only a few beamlines~\cite{SHINE:2018,Raubenheimer:2018,Liu:2023,Allaria:2013,Galayda:2018}.

A substantial gap thus remains in the parameter space between DLSRs and FELs: a source that combines pulse durations of sub-ps to 20~ps and moderate photon flux ($10^{6}$--$10^{11}$ photons per pulse), together with soft-to-hard x-rays and MHz-level repetition rates.
This operating  parameter regime is of considerable importance for a broad class of light source applications. 
Key experimental techniques that exploit it include split-pulse XPCS for studying equilibrium fluctuations in disordered materials~\cite{Roseker:2020}, serial
femtosecond crystallography and single-particle imaging of radiation-sensitive biological specimens~\cite{Chapman:2011,Seibert:2011,Fratalocchi:2011}, and
conformer-resolved spectroscopy and reaction dynamics of transient molecular species~\cite{Li:2024,Huang:2025}.  
Bridging this capability gap remains an open challenge and has motivated extensive research on storage-ring free-electron lasers (SRFELs), which aim to combine the complementary strengths of storage rings and FELs. 

 Early SRFELs, operating in the low-gain oscillator regime, have been demonstrated from the infrared to the ultraviolet~\cite{Billardon:1983,Couprie:1991}. 
Extending this approach to the Extreme Ultraviolet (EUV) and X-ray regions, however, has proven difficult owing to the stringent requirements on crystal Bragg reflectors and beam quality; related efforts remain under active investigation~\cite{XFELO,HuaYu}. 
In recent years, alternative storage-ring-based approaches have been explored. Coherent seeding schemes such as angular-dispersion microbunching, steady-state microbunching, and echo-enabled harmonic generation have been investigated~\cite{ADM,SSMB,xiaofancai,Liu:2026}. 
In some of these schemes, sub-ps longitudinally coherent pulses can be produced, but laser technology limits the light pulse repetition rate to \text{$\sim$1--10~kHz} \cite{DiMitri:2021}. 
 In parallel, the high-gain FEL regime has been pursued through the combined equilibrium scheme~\cite{Zhou:2023,Cai:2025}. 
 In this scheme, the  storage ring is combined with the FEL, and the electron beam then reaches a new steady state governed by the balance between FEL-induced heating and synchrotron radiation damping. 
 These approaches, however, face a fundamental tension between the FEL radiation and the repetition rate: high-gain lasing demands a high peak current, whereas high-repetition-rate operation requires the beam quality to be preserved over many turns. 
  As a concrete illustration, Cai and Wu~\cite{Cai:2025} showed that the FEL-induced energy spread suppresses the exponential gain, thereby reducing the radiation power growth to approximately linear with undulator length. 

It appears more viable to decouple the FEL interaction from the storage ring via a bypass configuration   
  \cite{DiMitri:2021,Feng:2017,Yang:2024}.
  The self-amplified spontaneous emission (SASE)   bypass proposed by Di Mitri \textit{et al.}\cite{DiMitri:2021} serves as a representative example, in which a bunch train is extracted from the DLSR, compressed in an RF cavity and magnetic arc to a peak current of $\sim$0.7 kA. The compressed bunch then traverses a transverse gradient undulator (TGU) for high-gain SASE radiation, after which it passes through a decompression section to restore its longitudinal phase space before reinjection into the storage ring.
Simulations indicate that this scheme delivers sub-ps FEL pulses with peak powers on the order of 100~MW in the EUV/soft-X-ray regime at an FEL repetition rate of 2.5 kHz. 
Nevertheless, increasing the FEL repetition rate toward the MHz level would require the stripline kicker to operate continuously at 227~kHz---a requirement far beyond the present state of the art.


Building on the bypass concept, what we are trying to develop is a distinctive storage ring-based light source concept called   dual-ring light source. In this scheme, an isochronous ring equipped with multiple TGUs is added to an existing DLSR, with the two rings connected by bunch compression (and decompression) systems and fast injection (and extraction) systems. 
The DLSR provides electron beams with MHz repetition rates, excellent stability, low transverse emittance, and multi-GeV energy; a downstream bunch compression system then shortens the bunch length to the ps scale. 
The key distinction from previous bypass schemes lies in the multi-turn circulation of compressed bunches in the isochronous ring. Bunches circulate for multiple turns (e.g., 24 turns), radiating at each pass through the TGUs.  
 This strategy decouples the effective FEL repetition rate from the kicker repetition rate, allowing the kicker to operate at a practical 10~kHz while the system delivers MHz-level FEL pulses from a single TGU. 
Relative to the DLSR alone, the proposed design increases the per-pulse photon number by 2--4 orders of magnitude.  
At the same time, it overcomes two principal drawbacks of linac-driven FELs, modest stability and limited beamline availability, while offering significant cost advantages over a dedicated superconducting FEL facility.
The dual-ring scheme thus fills  the parameter space between DLSRs and normal-conducting linac-driven FELs in repetition rate, pulse duration, and brilliance. It can simultaneously deliver  MHz FEL pulses of $\sim 6 \text{~ps}$ duration with $10^{7}$--$10^{9}$ photons per pulse across the 0.1--$4 \text{ nm}$ wavelength range.

 \section{Dual-ring light source concept}

\begin{figure}[!htb]
\includegraphics[width=0.8\hsize]{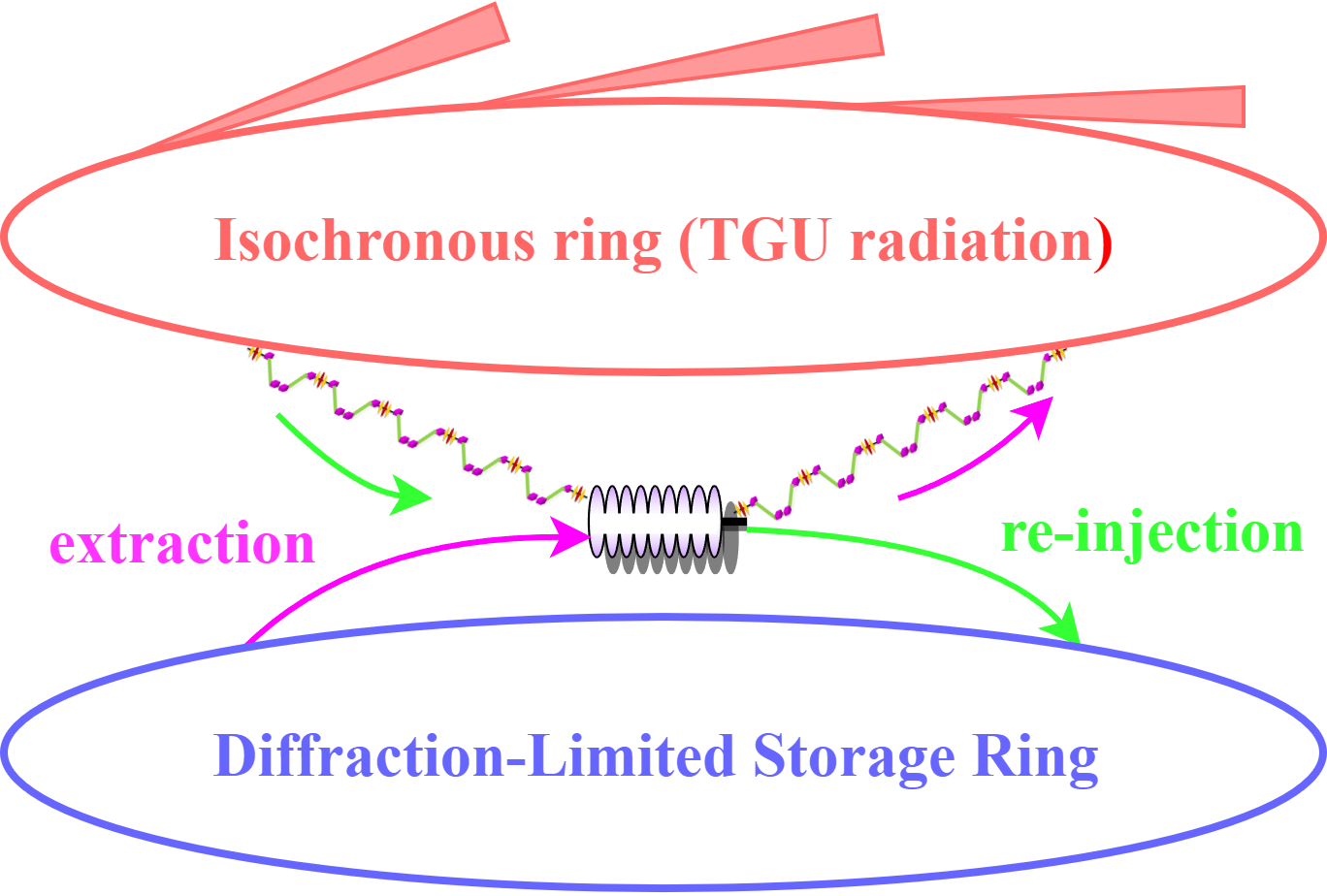}
\centering
\caption{Schematic layout of the dual-ring light source, including  a DLSR, an isochronous  ring, a bunch compression (and decompression) system, and a fast injection (and extraction) system.  }
\label{fig1}
\end{figure}

Figure~\ref{fig1} shows the layout of the dual-ring light source. 
The system
comprises a DLSR,   an isochronous ring equipped with
TGUs,  transport lines, and a fast injection and extraction system.

 The DLSR is based on a modified hybrid seven-bend achromat lattice~\cite{Jiao:2018},
with parameters similar to those of HEPS in high-brightness operation. 
 The main DLSR parameters are summarized in Table~\ref{tab:beam_params}. The ring stores
680 bunches at 6~GeV with an average current of 200~mA, and its calculated Touschek
lifetime is about 2~h~\cite{RDTM}. 
We assume a horizontal equilibrium emittance of 80~pm.rad and a coupling ratio of
$\varepsilon_y/\varepsilon_x = 2\%$. This emittance is at the same level as in
Ref.~\cite{Jiao:2018}, whereas the requirement of the present scheme is greatly relaxed.

\begin{table}[htbp]
\centering
\caption{DLSR main parameters }
\label{tab:beam_params}
\begin{tabular*}{8.5cm} {@{\extracolsep{\fill} } lcc}
  \toprule
Parameter  & Value & Units \\
\midrule
 Circumference length $C_{\mathrm{ring}}$ & 1360.4  & m\\
 Revolution period  $T_0$  & 4.5 & $\mu$s \\
Beam  energy $E_0$ & 6.0 & GeV \\
Damping time  $(x, y, z)$ & 10.2, 18.9, 16.4 & ms \\
Momentum compaction $ \alpha_c $  & $ 1.56\times10^{-5}$ &\\
Number of bunches $N_b$ & 680  & \\
 RF frequency  &　166.7　&　　MHz　\\ 
Bunch  charge  $Q$ & 1.33  & nC \\
Relative energy spread $\sigma_{\delta,0}$  & 0.11\% & — \\
Bunch length  $\sigma_{\text{z},0}$  & 1 & cm \\
Peak beam current $ I_{\text{pk}}$ & 16 & A\\
Average beam current $ I_{\text{average}}$ & 200 & mA \\
Touschek  lifetime &  2.0 &  h \\
Horizontal emittance  & 80 & pm.rad \\
Vertical emittance & 1.6 & pm.rad \\
 \bottomrule
\end{tabular*}
\end{table}

\begin{figure*}[!htb]
\centering
\includegraphics[width=0.9\hsize]{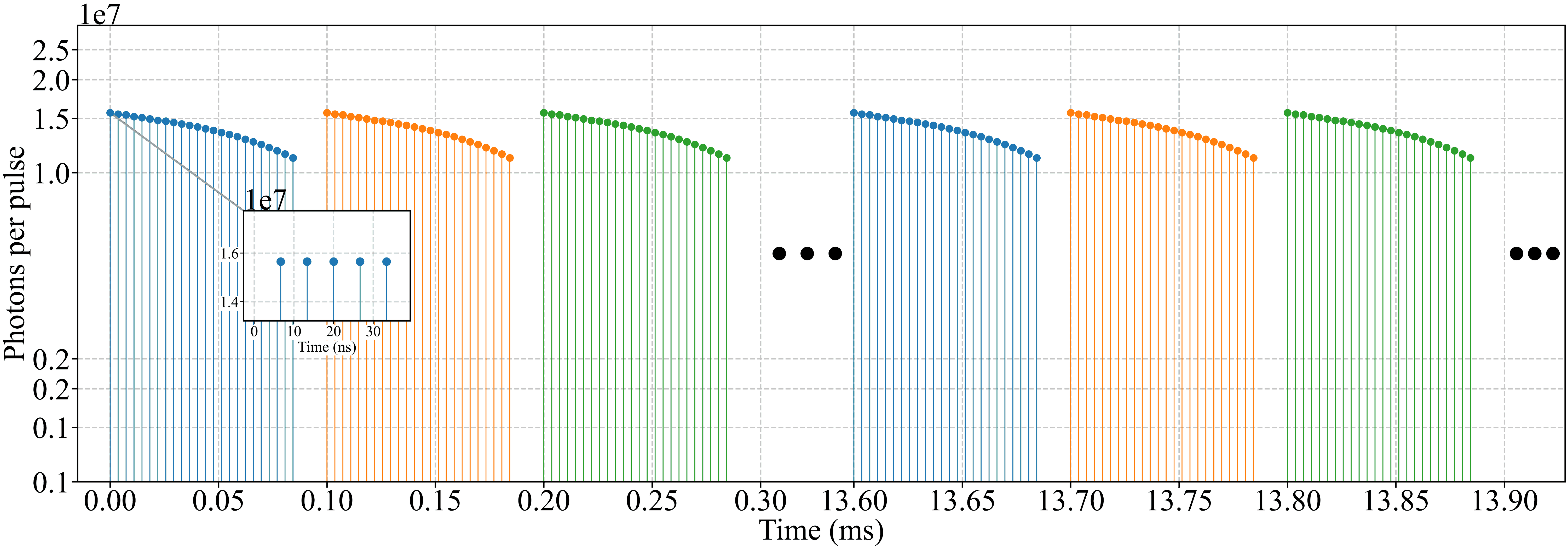}
\caption{ Photon number per pulse as a function of time over one complete extraction cycle, which
consists of 136 kicker periods (13.6~ms). The results are for 0.1~nm (12.4 keV) radiation from the
TGUs.  Only the first three batches (0--0.3~ms) are
displayed, in blue, orange, and green, respectively. }
\label{fig2xx}
\end{figure*}

The isochronous ring   consists of eight arc sections and eight straight sections. 
The arcs employ isochronous
three-bend achromat (TBA) lattices. Seven of the straight sections are occupied by TGUs, while the remaining one serves as the injection and extraction section.
 
The transport lines comprise three parts: a bunch compression system, a decompression system, and an $R_{56}$-free transport section. The $R_{56}$-free section is also built from TBA cells. The compression and decompression systems each consist of RF cavities and magnetic chicanes, and to reduce cost the same RF cavities are shared between the two stages.

In each cycle, a small fraction of the stored bunches (e.g., 5 out of 680) is extracted from the DLSR by the fast injection system and delivered to the bunch compression system. The compressed bunches are
injected into the isochronous ring, where they circulate for 24 turns and emit SASE
radiation in each TGU.  After lasing, the bunches are extracted from the isochronous ring, directed into the decompression system, and reinjected into the DLSR.  
The bunches of the reinjected batch undergo radiation damping in the DLSR, while the same
extraction--compression--radiation--decompression--damping sequence is applied to the next
batch. 
By the time all bunches have been processed in this way (136 batches in total), the first batch re-enters the cycle, as illustrated in Fig.~\ref{fig2xx}.

Note that the compression system is deliberately placed outside the isochronous ring. An in-ring scheme would offer a simpler dual-ring layout, with the RF cavities and 
chicanes occupying one or more straight sections. Nevertheless,  it  suffers from an
inherent drawback.  The
bunch would undergo a full compression--decompression cycle on every turn, leading to
cumulative emittance growth in both the transverse and longitudinal planes. This would
severely limit the effective number of radiation turns in the isochronous ring and thus
undermine the  advantage of the multi-turn radiation strategy.

To achieve an effective FEL repetition rate of $f_{\mathrm{FEL}} > 1$~MHz from a single
TGU,  
 we adopt a multi-turn radiation strategy  in which five bunches lase for  $ 24$ turns and are switched in and out by a fast kicker operating at a repetition rate of
$f_{k} = 10$~kHz. 
 Such a repetition rate is considered   feasible with current stripline kicker technology~\cite{DiMitri:2021}. 
The system employs multi-bunch operation, so that
\begin{equation}
f_{\mathrm{FEL}} \approx \frac{N_{\mathrm{iso}}}{T_{\mathrm{iso}}} > 1~\mathrm{MHz},
\end{equation}
where $T_{\mathrm{iso}} = C_{\mathrm{iso}}/c$ is the revolution period of the isochronous
ring, assumed to have a circumference of $C_{\mathrm{iso}} = 1.1$~km. 
With $N_{\mathrm{iso}} = 5$, the scheme yields $f_{\mathrm{FEL}} \approx 1.36$~MHz. 
The kicker repetition rate also bounds the number of turns each extracted bunch can spend in the isochronous ring: at
most $1/(T_{\mathrm{iso}} f_{k}) = 27.3$~turns. After accounting for the time overhead of
extraction and injection, we conservatively set the effective number of lasing turns to
24.     The
parameters of the dual-ring coupling system are summarized in Table~\ref{tab:iso_params}.

\begin{table}[htbp]
\centering
\caption{ Related parameters  after introducing the isochronous ring   }
\label{tab:iso_params}
\begin{tabular*}{8cm} {@{\extracolsep{\fill} } lc}
\toprule
Parameter    & Value \\
 \midrule
Kicker repetition rate &     10 kHz\\
Isochronous ring circumference & 1.1 km\\
Turns in isochronous ring & 24 \\
Number of  extracted  bunches  & 5 \\
Effective FEL repetition rate & $\approx$1.36 MHz \\ 
Damping turns in the DLSR & $\approx$2978 turns \\
 \bottomrule
\end{tabular*}
\end{table}

The bunches reinjected into the DLSR undergo sufficient radiation damping to largely
recover their beam parameters before their next extraction.   
Extracting five bunches per cycle requires 136 extraction cycles to process all
$N_{b} = 680$ bunches stored in the DLSR. 
From its reinjection until its next extraction,
the first batch therefore waits $N_{b}/N_{\mathrm{iso}} - 1 = 135$ kicker periods and
accumulates
\begin{equation}
\frac{1}{T_{0} f_{k}}\left(\frac{N_{b}}{N_{\mathrm{iso}}} - 1\right) = 2978
\end{equation}
turns of radiation damping in the DLSR, where $T_{0}$ is the revolution period of the
DLSR. The transverse and longitudinal damping times, expressed in turns, are
$(n_{x}, n_{y}, n_{z}) = (2249, 4168, 3617)$,  comparable to the 2978-turn interval between successive extractions of the same batch, so that much of the beam quality is restored in each cycle. Under the combined
action of radiation damping in DLSR and the persistent FEL perturbation in the isochronous ring,
the beam reaches a new equilibrium.










\begin{figure}
\centering
\includegraphics[width=0.45\textwidth]{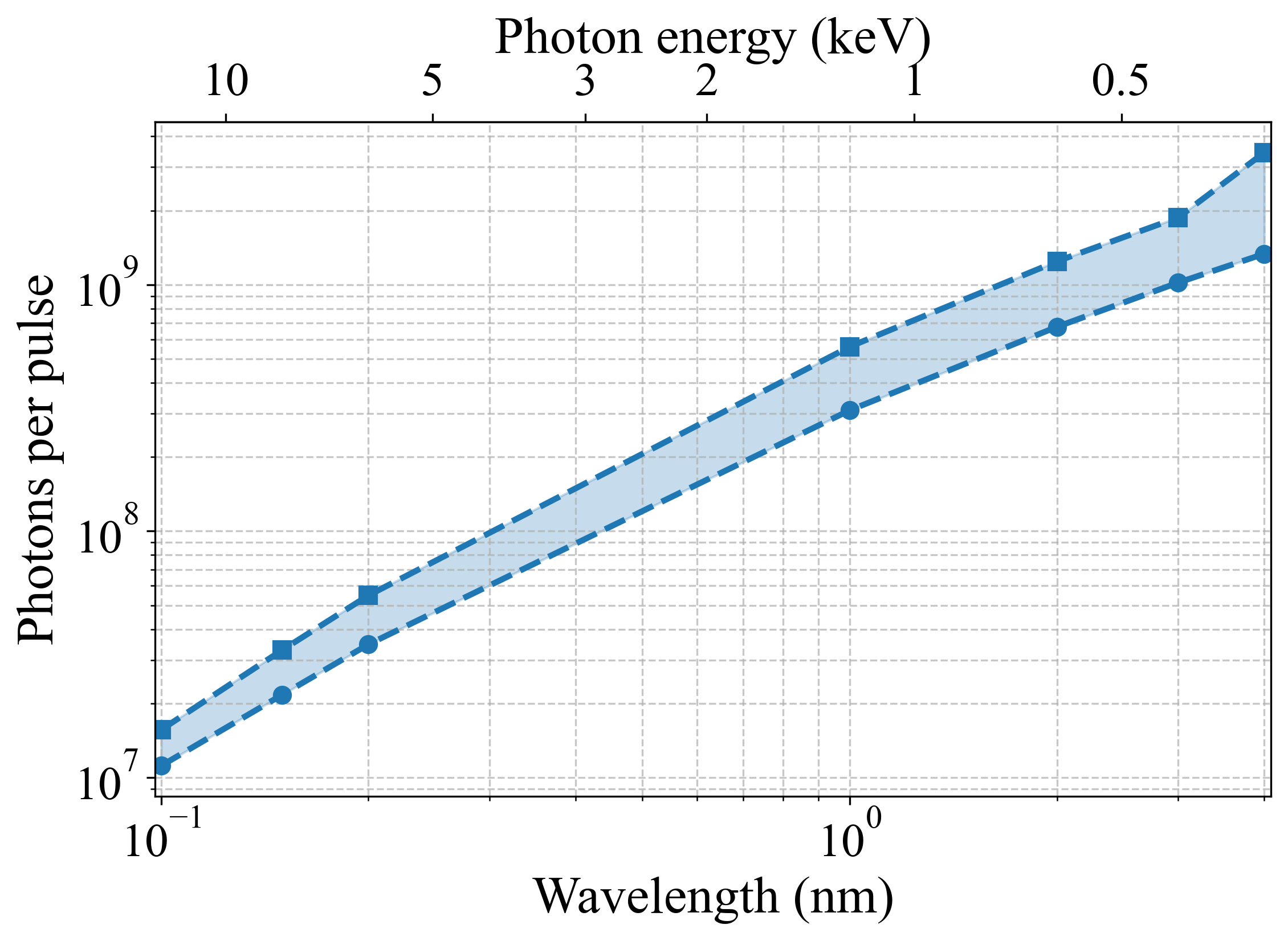}
\caption{Achievable photon number per pulse vs wavelengths of \text{0.1~nm}, 0.15 nm, 0.2 nm, 1 nm, 2 nm, 3 nm, and 4 nm. The bar indicates the range over the 24 lasing turns.  }
\label{fig3}
\end{figure}
  
  The dual-ring design also delivers a high photon number per pulse. At 0.1 nm, the scheme yields $1.12 \times 10^{7}$--$1.56 \times 10^{7}$ photons per pulse; results for other hard and soft X-ray wavelengths are shown in Fig. \ref{fig3}. Achieving this performance requires precise control of the beam quality, as discussed in the following sections.

   The preceding analysis has taken a 6 GeV HEPS-type storage ring as the DLSR, but the dual-ring concept is not specific to this energy and applies
to other energies as well, with several clear changes. 
Lowering the beam
energy by a factor $r$, the circumferences of both rings shrink by $r$;  in compressor system, 
the  cavity voltage drops by $r$ and the chicane shortens according
to $L_{\text B}^2 L_{\text{d1}} \propto r^2$, where $L_{\text B}, L_{\text{d1}}$  the  length of bend and first drift. 
For $r \approx 10$, the chirp voltage is
reduced to $\sim$10 MV, which brings it within the reach of
normal-conducting RF and thereby lowers the cost.
However, lowering the
beam energy also brings drawbacks: the CSR-induced energy spread scales as
$\gamma^{-1}$ and worsens by $r$, while for the radiation the photon energy
drops by $r^2$. Of course, a concept design at another energy would require
more accurate calculations and would face additional issues, which lie
beyond the scope of this paper.

\section{Lattice design of the isochronous ring}\label{sec:lattice}

The isochronous ring contains   40 identical TBA cells organized into eight arcs of five cells each. This section details the design of the TBA cell.

\subsection{TBA cell  }\label{sec:TBA}

\begin{figure}[!h]
\centering
\includegraphics[width=0.45\textwidth]{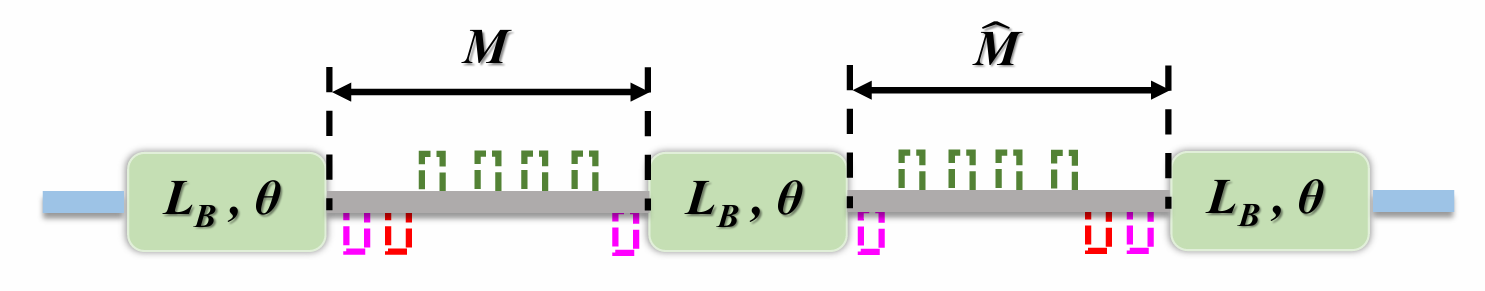}
\caption{Schematic of a TBA cell. The lattice section between dipoles containing, e.g., quadrupoles, sextupoles, and octupole.    }
\label{fig:TBA}
\end{figure}

The TBA cell  follows a symmetric layout. Its   It contains   three identical dipoles, each 1.8 m long with a bending angle of $3^{\circ}$. Four quadrupole families, two sextupole families, and one octupole family are inserted between the dipoles, as shown in \text{Fig. \ref{fig:TBA}}.

The TBA cell is optimized for two objectives: i) minimizing the normalized transverse emittance growth $\Delta\varepsilon$ induced by a single cell (in both the horizontal and vertical planes), and ii) controlling the longitudinal dispersion terms $R_{56}$, $T_{566}$, and $U_{5666}$. Since the bunches circulate for 24 turns, the single-cell optimization is validated over the full path to ensure that the cumulative emittance growth and bunch-length distortion remain acceptable.

 The four quadrupole families each have a fixed length of 0.4 m. The gradients of two families are  free parameters, while the remaining two are constrained by the achromatic conditions ($R_{16}=0$ m, $R_{26}=0$) and the isochronous condition ($R_{56}=0$ m). The eight drift lengths, the two sextupole strengths, and the octupole strength are also treated as free variables, within the following ranges:
\begin{equation}
 \begin{aligned}
K \in [-5, 5]\ \mathrm{m^{-2}},&~~~~
S \in [-300, 300]\ \mathrm{m^{-3}}, \\
O \in [-6000, 6000]\ \mathrm{m^{-4}},&~~~~
L_{d} \in [0.1, 4]\ \mathrm{m}.
\end{aligned}
\end{equation}


The optimization uses the multi-objective particle swarm optimization (MOPSO) algorithm~\cite{MOPSO}, which is well suited for global search in high-dimensional parameter spaces. Particle tracking is performed with \textsc{Elegant} \cite{elegant}, including higher-order transport matrices, transient and steady-state CSR in the dipoles, CSR in the drift spaces, and ISR. Note that the influence of chromaticity-driven instabilities on the beam transport is not taken
into account in the present study. Since the growth times of single-bunch instabilities are
typically several thousand turns, whereas each bunch circulates in the isochronous ring for
only 24 turns, the growth accumulated over this interval is very limited.

The following constraints guide the optimization: optical stability, i.e., $|m_{11}+m_{22}| < 2$ and $|m_{33}+m_{44}| < 2$; horizontal $\beta$-functions in the range $(0.2 \text{ m}, 100\text{ m})$; a total cell length not exceeding 16 m; and residual dispersion $|R_{16}| < 10^{-6}$~m, $|R_{26}| < 10^{-6}$. Any design violating a constraint is penalized by multiplying both objective functions by a factor greater than 1.

\begin{figure}
\centering
\includegraphics[width=0.4\textwidth]{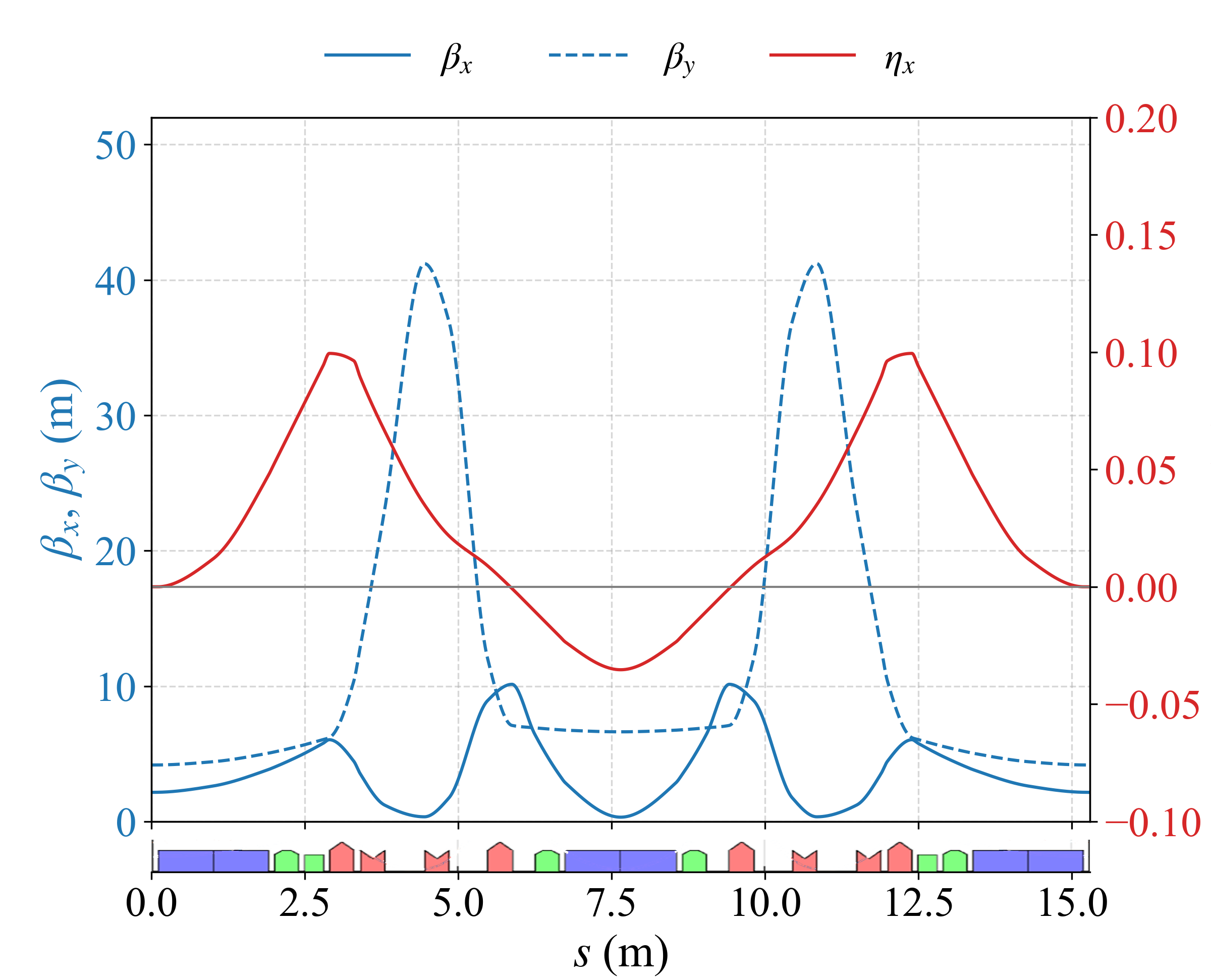}
\caption{Betatron and dispersion functions of a single TBA cell.      }
\label{fig:twiss}
\end{figure}

After convergence, the algorithm yields a set of solutions distributed along the Pareto front, trading off emittance preservation against isochronicity. From these candidates, we select a design that balances low exit emittance with low exit bunch-length. The TBA cell parameters are summarized in Table~\ref{tab:tba_opt}, and the linear optics functions are shown in Fig.~\ref{fig:twiss}.

\begin{table}[h]
\centering
\caption{Optimized parameters of a single TBA cell.}
\label{tab:tba_opt}
\begin{tabular*}{7cm} {@{\extracolsep{\fill} } lc}
\toprule
Parameter   & Value \\
\midrule 
\quad Q1 strength   & \phantom{$-$}2.9396\,m$^{-2}$ \\
\quad Q2 strength   & $-0.6165$\,m$^{-2}$ \\
\quad Q3 strength   & $-2.5699$\,m$^{-2}$ \\
\quad Q4 strength   & \phantom{$-$}3.4286\,m$^{-2}$ \\  
\quad S1 strength   & \phantom{$-$}213.5687\,m$^{-3}$ \\
\quad S2 strength  & \phantom{$-$}299.7176\,m$^{-3}$ \\
\quad O1 strength & $-6000$\,m$^{-4}$ \\ 
\quad $D_{1}$  & 0.10\,m \\
\quad $D_{2}$  & 0.10\,m \\
\quad $D_{3}$  & 0.10\,m \\
\quad $D_{4}$  & 0.10\,m \\
\quad $D_{5}$  & 0.65\,m \\
\quad $D_{6}$  & 0.63\,m \\
\quad $D_{7}$  & 0.37\,m \\
\quad $D_{8}$  & 0.10\,m \\
\midrule 
\quad Horizontal tune    & $63.3 $\\
\quad Vertical tune   & $13.5 $ \\
\quad   $\max\beta_{x}$ & 10.13\,m \\
\quad   $\max\beta_{y}$ & 41.20\,m \\
\quad   $\max\eta$ & 0.1\,m \\ 
\quad TBA cell length   & 15.3\,m \\
\quad TBA cell $R_{56}$  & $9.78 \times 10^{-9}$\,m \\
\bottomrule
\end{tabular*}
\end{table}

\begin{figure}[!h]
\centering
\includegraphics[width=0.38\textwidth]{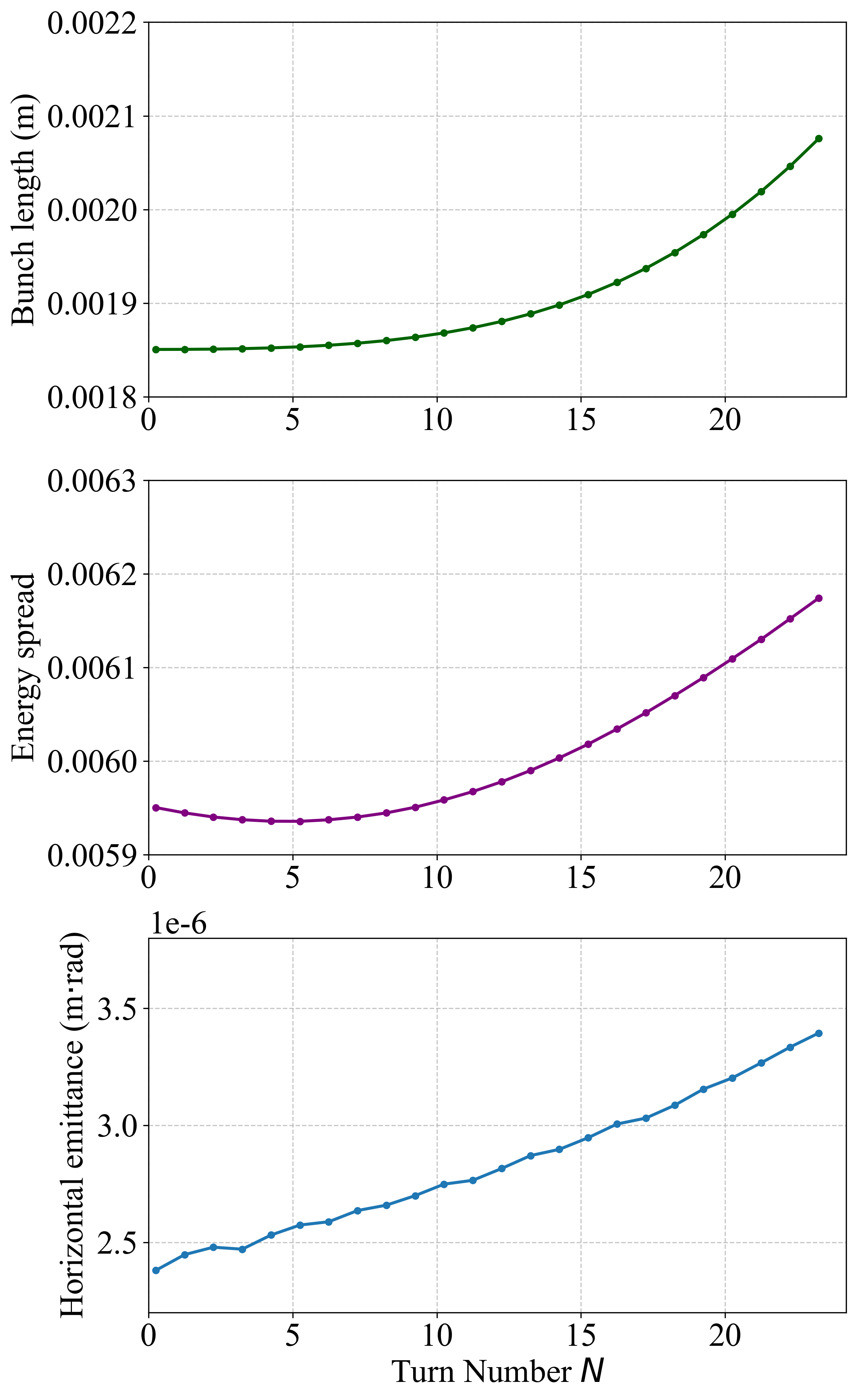}
\caption{The evolution of the beam quality, the bunch length, energy spread, and  normalized emittance, with turn number in the isochronous ring.     }
\label{fig:turn}
\end{figure}

\begin{figure}[!h]
\centering
\includegraphics[width=0.5\textwidth]{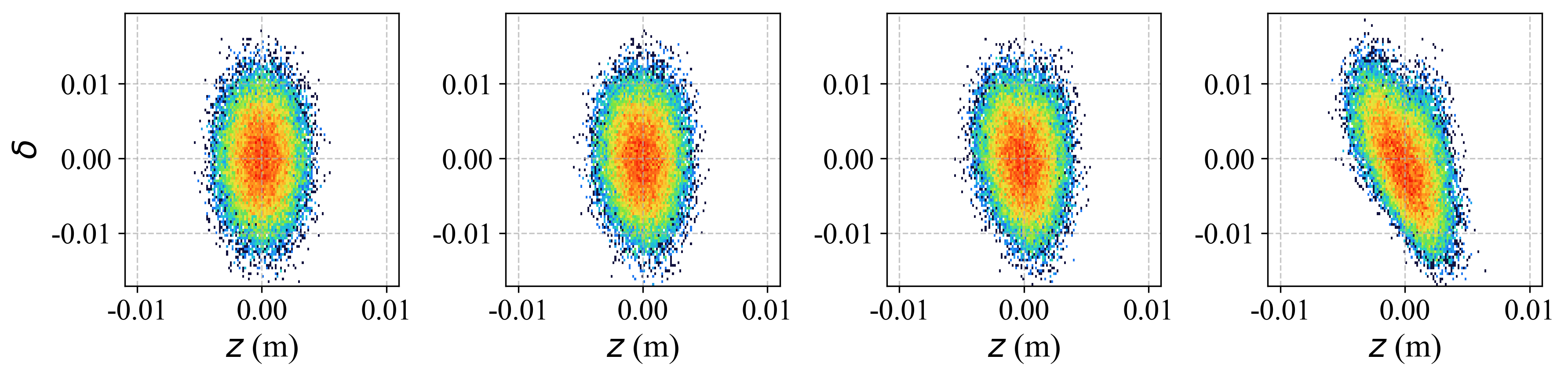} 
\caption{Longitudinal phase space at the 1st, 5th, 10th, and 24th turn in the isochronous ring.   }
\label{t_p_density_4panels}
\end{figure}

We performed particle tracking to characterize the multi-turn beam transport in the
isochronous ring. The initial bunch parameters are taken from the compression results
(see Table \ref{tab:chicane_params}).  The evolution of the beam parameters with turn number is shown in \text{Fig.~\ref{fig:turn}}, and the longitudinal phase-space distributions at selected turns are plotted in Fig. \ref{t_p_density_4panels}. Over the 24-turn cycle, the bunch length, energy spread, and  normalized emittance increase by 13.7\%, 4.1\%, and 47.3\%, respectively. This corresponds to an average growth of less than 2\% per turn, confirming that the beam quality is well preserved over multiple turns.

\begin{figure}[!h]
\centering
\includegraphics[width=0.45\textwidth]{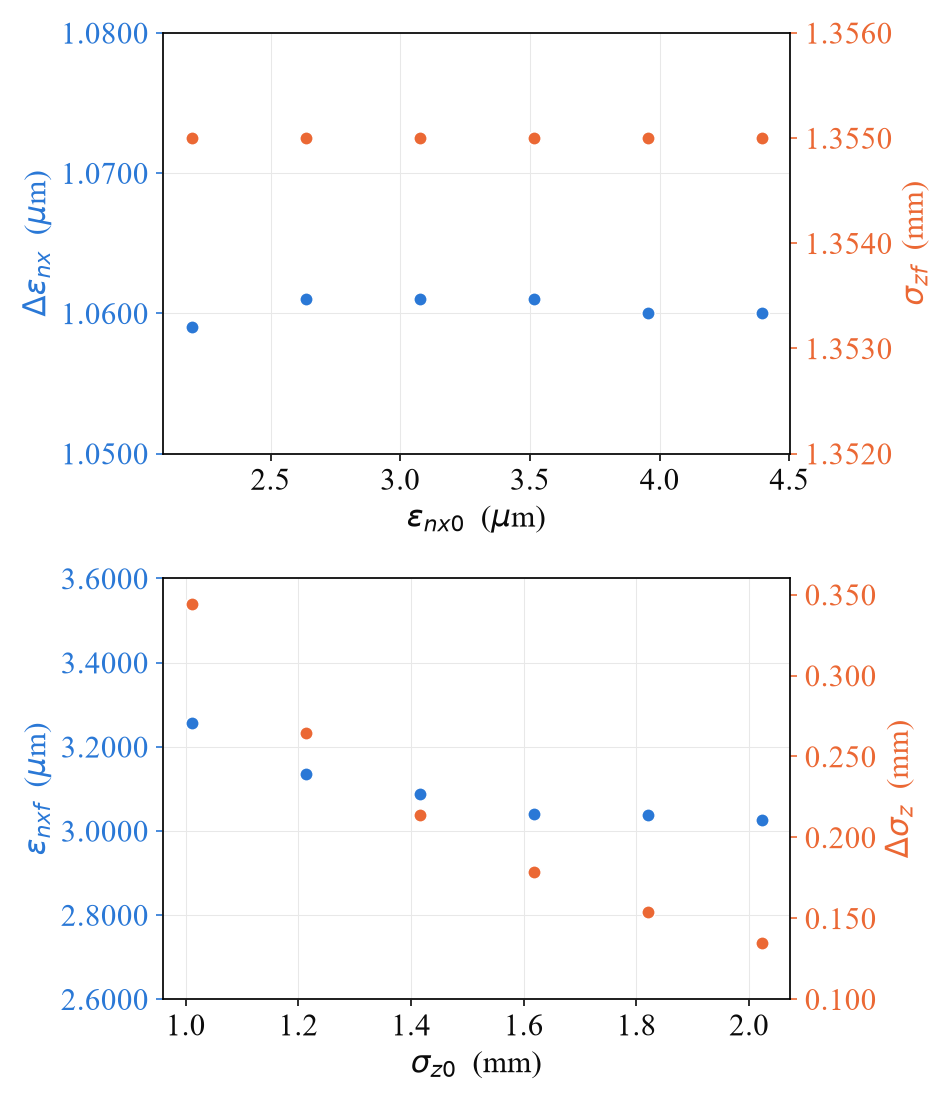} 
\caption{	The bunch length and  normalized emittance growth after 24 turns in  isochronous ring over a range of    initial   bunch lengths  and  normalized emittances.    }
\label{scan_comparison}
\end{figure}

In the dual-ring scheme, the bunch lengthens and the transverse emittance grows gradually during the multi-turn circulation in the isochronous ring. We therefore evaluate the robustness of the lattice against variations in the initial bunch length and normalized emittance. As shown in Fig.~\ref{scan_comparison}, the multi-turn transport is insensitive to the initial emittance, while longer bunches experience less degradation of the beam quality. These tests indicate that the proposed lattice is robust and remains stable throughout the multi-turn process.

\subsection{ transverse gradient undulator   }\label{sec:straight}
 
The isochronous ring consists of eight arc sections interleaved with eight straight sections of equal length. Seven of the straight sections house TGUs for FEL radiation; the remaining one accommodates the injection and extraction system. Optical matching is performed between each arc and the adjacent undulator.

 The compression shortens the bunch duration and leaves a correspondingly larger
energy spread of $\sigma_{\delta, \text{f}} \sim 0.6\%$, which exceeds the typical FEL gain bandwidth ($\rho \lesssim 0.1\%$).  To preserve the FEL gain, this large energy spread is compensated by using TGUs~\cite{Huang:2012}.   When the matching condition
\begin{equation}\label{eq:TGU}
\alpha D = \frac{2 + K_0^2}{K_0^2}
\end{equation}
is satisfied, the energy-dependent term in the FEL resonance condition is compensated. Here $\alpha$ is the transverse gradient parameter, $D$ is the vertical dispersion, and $K_0$ is the undulator parameter.  
The TGU is installed in a straight section with a
prescribed vertical dispersion, which is generated by a dogleg at the TGU entrance.


The FEL fundamental wavelength of emission is specified by the resonance condition
\begin{equation}
\lambda_s = \frac{\lambda_u}{2\gamma^2} \left(1+\frac{K_0^2}{2}\right),
\label{resonance}
\end{equation}
where $\lambda_u$ is the undulator period. Here $K_0 $ is related to $\lambda_u$ and undulator peak magnetic field $B_{\text{peak}}$.    Assuming the relation between $B_{\text{peak}}$
and  the gap $g$ is given by the Halbach formula  for rare earth cobalt magnets, we have \cite{Halbach:1983}
\begin{equation}
K_0 = \frac{e\lambda_u B_{\text{peak}}}{2\pi mc} = 93.43\,\lambda_u[\text{m}]\,B_{\text{peak}}[\text{T}],
\end{equation}
\begin{equation}
B_{\text{peak}} = 3.33 \exp\left(-5.47 \frac{g}{\lambda_u} + 1.8\left(\frac{g}{\lambda_u}\right)^2\right).
\end{equation}
 
The TGU parameters are chosen to cover the target wavelength range from 0.1 nm (hard X-ray) to 4 nm (soft X-ray). A peak magnetic field of $B_{\mathrm{peak}}=0.8$ T is assumed, a value readily achievable with permanent-magnet undulators.  Table \ref{tab:tgu_params} summarizes the undulator parameters at selected wavelengths.   
The undulator length is set to $L_u = 50$~m, as dictated by the FEL
performance discussed in Sec.~\ref{sec:photon}; within the $61$~m-long
straight sections this leaves about $11$~m for the vertical doglegs and the
Twiss matching sections at the two ends of the TGU.

\begin{table}[htbp]
\centering
\caption{TGU parameters for selected FEL wavelengths. The peak field is fixed at $B_{\text{peak}}=0.8$~T.   }
\label{tab:tgu_params}
\begin{tabular*}{8.8cm} {@{\extracolsep{\fill} } c c c c c c c c}
\toprule
$\lambda_s$ & 0.1 nm & 0.15 nm & 0.2 nm & 1 nm & 2 nm & 3 nm & 4 nm \\
\midrule
$\lambda_u$ (cm) & 1.60 & 1.98 & 2.26 & 4.36 & 5.62 & 6.49 & 7.17 \\
$K$ & 1.20 & 1.48 & 1.69 & 3.26 & 4.20 & 4.85 & 5.36 \\
$g$ (cm) & 0.46 & 0.57 & 0.65 & 1.26 & 1.62 & 1.87 & 2.06 \\
$D$ (cm) & 4.95 & 4.15 & 3.64 & 2.02 & 1.73 & 1.52 & 1.42 \\
$\alpha ~(\text{m}^{-1})$  &48.3   & 46.2 & 46.6 &  58.8 &64.4 &  71.4 &75.3\\
\bottomrule
\end{tabular*}
\end{table}
 
 The vertical dispersion $D$ is optimized for maximum peak power using Xie's fitting formula~\cite{Xie:1995}; the optimum values are listed in Table~\ref{tab:tgu_params}. The corresponding transverse gradient parameter $\alpha$ is obtained from Eq. \ref{eq:TGU}. The required gradient ranges from 46.2 m$^{-1}$ to 75.3 m$^{-1}$, which is within the capability of current technology \cite{Huang:2012}.
    
     Vertical doglegs on either side of the TGU provide the required dispersion
\(D=\theta(2L_\text{B}+L_\text{d})\). At 0.1 nm, with \(\theta=0.02\) rad, \(L_\text{B}=0.4\) m,
and \(L_\text{d}=1.675\) m, \(D=4.95\) cm; for other wavelengths, \(\theta\) is
scaled in proportion to the required $ D $.  Twiss matching section is
inserted between each dogleg and the adjacent TBA.

 \section{Physics design of the transport line}
 The transport line consists of an isochronous TBA section and a chicane-based bunch compressor
and decompressor. The isochronous TBAs deflect the beam extracted from the DLSR and deliver it
to the compressor while preserving the beam quality. The TBA cell design is identical to that
used in the isochronous ring, so that the transport line reuses the ring's
standard arc module without additional design effort 
(see Sec.~\ref{sec:TBA}). 

 A decompressor is also required, but because it shares the same principle and design as the
compressor, the following discussion focuses mainly on the compressor. Section~\ref{sec:basic}
introduces the basic physics of bunch compression. Section~\ref{sec:challenge} discusses the principal effects that complicate the chicane design, namely coherent and incoherent synchrotron radiation (CSR and ISR). 
Section~\ref{sec:detailed} then presents a specific chicane design and evaluates its compression
performance and beam quality preservation.


\subsection{Basic physics of the compressor}\label{sec:basic}
The principle of  bunch compression is analogous to  chirped pulse amplification technique in the laser community \cite{Strickland:1985}. An electron beam is imprinted with an energy chirp by an RF cavity, and then passes through a longitudinal dispersive section, where particles with higher (lower) energy at the tail (head) of the bunch travel a shorter (longer) path \cite{Bartosik:2022}. 
  The RF cavity is operated at the zero-crossing phase $\phi_{\mathrm{rf}} = 0$, imparting a linear energy chirp $h = dE/(E dz)$ to the bunch. The subsequent dispersive section converts this energy chirp into a reduction in bunch length via its longitudinal dispersion $R_{56}$. The rms bunch length after compression is given by
\begin{equation}
\sigma_{\text{z,f}}^{2} = \sigma_{\text{z,0} }^{2} (1 + h R_{56})^{2} + (R_{56} \sigma_{\delta,0})^{2},
\end{equation}
where $\sigma_{\text{z,0}}$ and $\sigma_{\delta,0}$ are the rms bunch length and relative uncorrelated energy spread before compression, respectively. 
In the case of full compression ($h = -1/R_{56}$), the first term vanishes and the final bunch length is determined entirely by the second term: 
\begin{equation}\label{eq:6}
\sigma_{\text{z,f}} =| R_{56} \sigma_{\delta,0}|,
\end{equation}
corresponding to a compression factor $C = \sigma_{\text{z,0}}/|R_{56} \sigma_{\delta,0}|$.

\begin{figure}[!h]
\centering
\includegraphics[width=0.45\textwidth]{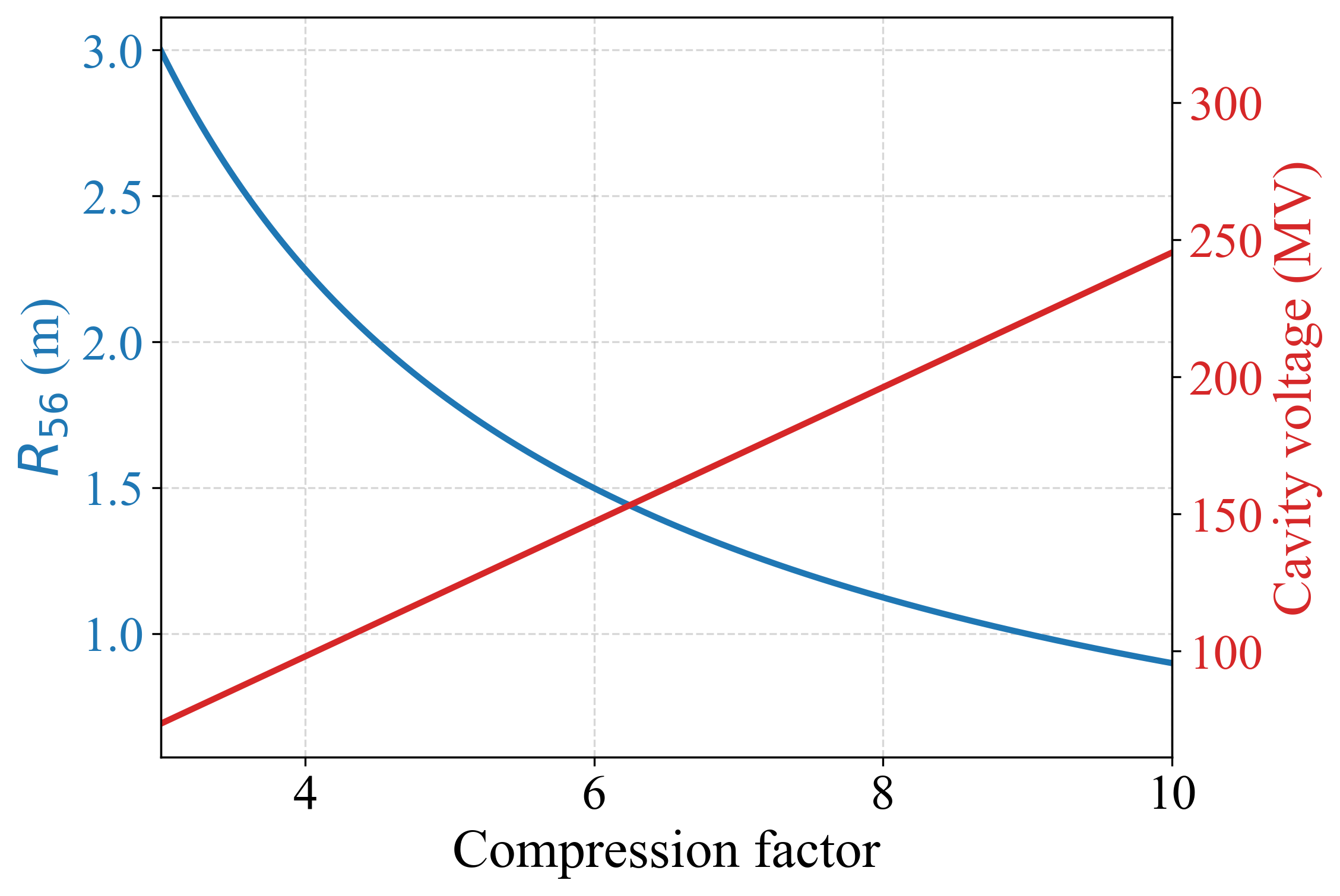}
\caption{Required $R_{56}$ and RF cavity voltage $V_{\mathrm{rf}}$ as functions of the compression factor $C$. The RF cavity is operated at 1.3 GHz. }
\label{trade-off}
\end{figure}

 Figure~\ref{trade-off} shows the required $R_{56}$ and RF cavity voltage $V_{\mathrm{rf}}$ as functions of the compression factor $C$. The two curves reveal an inherent trade-off. A larger compression factor demands a smaller $R_{56}$ but a higher $V_{\mathrm{rf}}$, which increases the cost of the superconducting RF system. The compression factor is therefore kept as low as the FEL requirements allow, so that the cavity voltage remains economically feasible.

A chicane is adopted for the magnetic compression stage. Its longitudinal dispersion is
\begin{equation}R_{56} = 2  \theta^{2} \left(L_{\text{d1}} + \frac{2}{3} L_{\text{B}}\right),
\end{equation}
where $L_{\text{B}}$, $L_{\text{d1}}$, and $\theta$ denote the bend magnet length, drift length, and bend angle, respectively. By contrast, a DBA-based arc compressor yields only $R_{56} = L_{\text{B}} \theta^{2}/3$. For the same required $R_{56}$, a chicane thus demands shorter or weaker bends, which helps mitigate beam quality degradation during compression, as discussed in Sec.~\ref{sec:challenge}.

\subsection{Design challenges: CSR and ISR effects}\label{sec:challenge}
The ability to compress an electron bunch with minimal degradation of beam quality is critical for the dual-ring design. 
However, when bent,   electrons    emit broadband synchrotron radiation, leading to CSR and ISR effects.  If not well controlled, these effects can severely degrade the beam quality, producing additional energy spread and projected transverse emittance   dilution.
Assuming a steady-state and a Gaussian beam distribution in the dipole, the CSR-induced energy deviation $\Delta \delta_{\mathrm{CSR}}$ is~\cite{Hajima}
\begin{equation}
\Delta \delta_{\mathrm{CSR}} = 0.2459 \frac{N_{p} r_{e}}{\gamma \sigma_{\text{z,0}}^{4/3}} \rho^{1/3} \theta,
\end{equation}
here $N_p$ is the electron population, $r_e$ is the classical electron radius, and $\gamma$ is the relativistic Lorentz factor. 
 The CSR-induced emittance growth at the chicane exit can be expressed in terms of the net CSR kick   $(\Delta x_{\mathrm{CSR,f}}, \Delta x'_{\mathrm{CSR,f}})$ as  
\begin{equation} 
  \varepsilon_f^{2} = \varepsilon_{0}^{2} + \varepsilon_{0}\Bigl(\beta_{\text{xf}} \langle\Delta x_{\mathrm{CSR,f}}^{\prime 2}\rangle   + 2\alpha_{\text{xf}} \langle\Delta x_{\mathrm{CSR,f}},\Delta x_{\mathrm{CSR,f}}'\rangle + \gamma_{\text{xf}} \langle\Delta x_{\mathrm{CSR,f}}^{2}\rangle\Bigr), 
\end{equation}
where $\varepsilon_{0}$ is the initial normalized emittance, and $\alpha_{\text{xf}}$, $\beta_{\text{xf}}$, $\gamma_{\text{xf}}$ are the Courant-Snyder parameters at the chicane exit.
The net CSR kick is obtained by accumulating the point-kick contributions from each bending magnet along the chicane~\cite{Zeng:2025,Zeng:20252}.

 The ISR-induced energy deviation and relative emittance growth, arising from incoherent quantum excitation in the bending magnets, can be estimated as \cite{Chao:2006,EEHG2}
 \begin{equation}
 \begin{aligned}
\Delta \sigma^2_{E,\text{ISR}} &= \frac{55e^2 \hbar c}{48\sqrt{3}} \frac{L_{\text{B} }}{\rho^3} \gamma^7, \\
\frac{\Delta\varepsilon}{\varepsilon_{0}} &\approx a E^{6} \frac{|\theta|^{5}}{\varepsilon_{0} L_{\text{B}}^{2}}\left(  L_{\text{d1}} + L_{\text{B}} + \frac{\beta_{\min} + \beta_{\max}}{2}\right),
\end{aligned}\label{ISR}
\end{equation}
with $a = 8 \times 10^{-8} ~ \mathrm{m^{2} \cdot  GeV^{-6}}$,    $\beta_{\min}$, $\beta_{\max}$ the minimum and maximum beta functions in the compression process.

\begin{figure}
\centering
\includegraphics[width=0.45\textwidth]{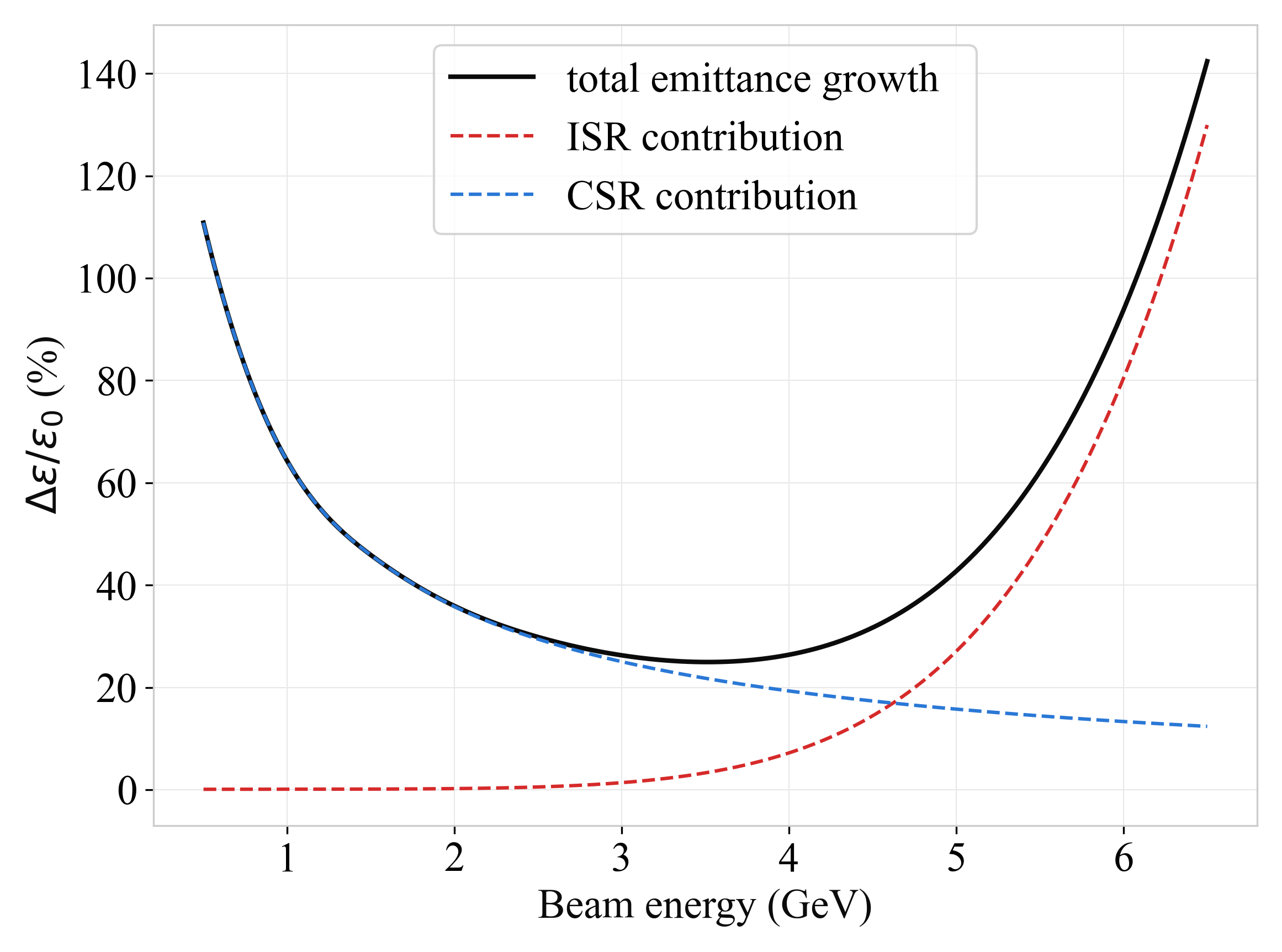}
\caption{CSR  and ISR   contributions to normalized emittance   growth for different beam energy using a $R_{56} = -0.5$ m chicane ($L_{\text{B}} = 3.5$ m, $L_{\text{d1}} = 7.5$ m, $L_{\text{d2}} = 1$ m, $\theta = 0.159$ rad, total length 30 m).}
\label{CSR_ISR_contribution}
\end{figure}

A key factor governing the relative importance of CSR and ISR is the beam energy. To illustrate this, we consider a symmetric C-chicane with $R_{56} = -0.5$ m, corresponding to a compression factor of $C = 12.5$. The beam energy is scanned from 0.5 GeV to 6.5 GeV, with the RF voltage scaled proportionally to keep the initial linear energy chirp constant. We employ a widely-used particle tracking code  \textsc{Elegant}  \cite{elegant}.  The results are shown in Fig. \ref{CSR_ISR_contribution}. At low energies, CSR dominates the relative emittance growth, and its contribution gradually decreases with increasing energy. Beyond $\sim$3 GeV, ISR becomes appreciable and eventually overtakes CSR as the dominant mechanism above 4.6 GeV.

Both effects impose constraints on the chicane design. The CSR contribution scales with the bunch peak current and CSR effects can be suppressed through an appropriate choice of chicane geometry. Specifically, a four-bend non-symmetric S-shaped layout can achieve CSR effect cancellation, and its beam quality preservation is demonstrated in Ref. \cite{Zeng:2025}.  The ISR contribution  dominates at high beam energies. Unlike CSR effects, ISR effects cannot be   cancelled. However,  they can be effectively mitigated by reducing the bending angle, since $\Delta\varepsilon/\varepsilon_{0} \propto |\theta|^{5}$ (see Eq.~\ref{ISR}).


\subsection{A detailed chicane design}\label{sec:detailed}
\begin{figure}
\centering
\includegraphics[width=0.32\textwidth]{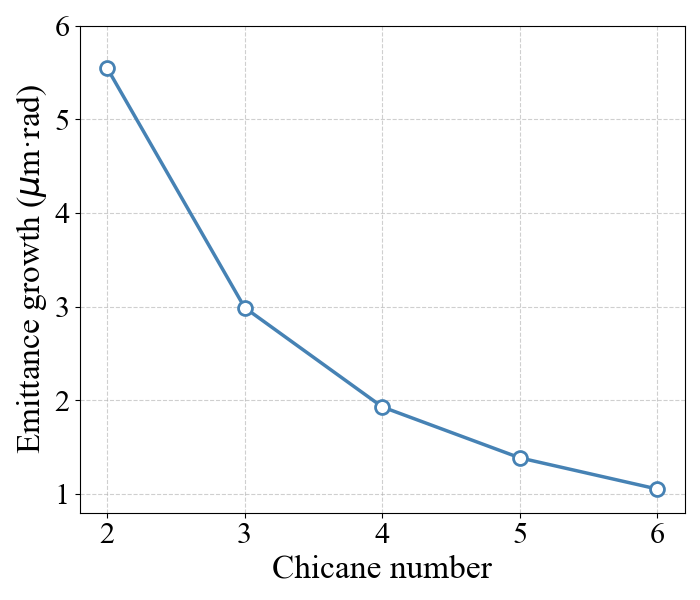} 
\caption{	Normalized emittance growth of chicane compressor vs the number of chicanes. A total $R_{56} = -1.63$ m (corresponding to $C = 5.5$) is distributed over the chicanes, each with a bend length of $L_{\text{B}} = 3.0 \text{ m}$, a drift length of $L_{\text{d1}} = 9$ m, and a total length of \text{30 ~m}. }
\label{chicane_number_comparison}
\end{figure}

In our case study, the beam energy is 6 GeV, where ISR effects dominate  the  normalized emittance growth. To mitigate ISR, we distribute the total required $R_{56}$ across multiple chicanes in series, reducing the bending angle per chicane. As shown in Fig. \ref{chicane_number_comparison}, increasing the number of chicanes suppresses   emittance growth. Balancing the physical footprint against emittance reduction, we adopt five chicanes in our compressor design.

For the full-compression scenario considered here, the minimum bunch length is set by the
initial slice energy spread and $R_{56}$ (see Eq.~\ref{eq:6}).
 The shorter the compressed
bunch, the larger the compression factor $C$, which raises the peak current and thereby
enhances the FEL performance. 
An excessively short bunch is, however, more susceptible to degradation during the long transport through the isochronous ring, as discussed in
Sec.~\ref{sec:photon}. 
  Balancing these competing demands yields a compression factor of $C = 5.5$, corresponding to a total RF cavity voltage of 135 MV and a total $R_{56} = -1.63$~m. 
	The RF cavity voltage can be provided by eight 9-cell 1.3 GHz cavities  (one cryomodule with \text{16.9 MV/m}), and the total  $R_{56}$  is distributed among five identical chicanes, each contributing  $ -0.326~\text{m}$.

\begin{figure}[!h]
\centering
\includegraphics[width=0.32\textwidth]{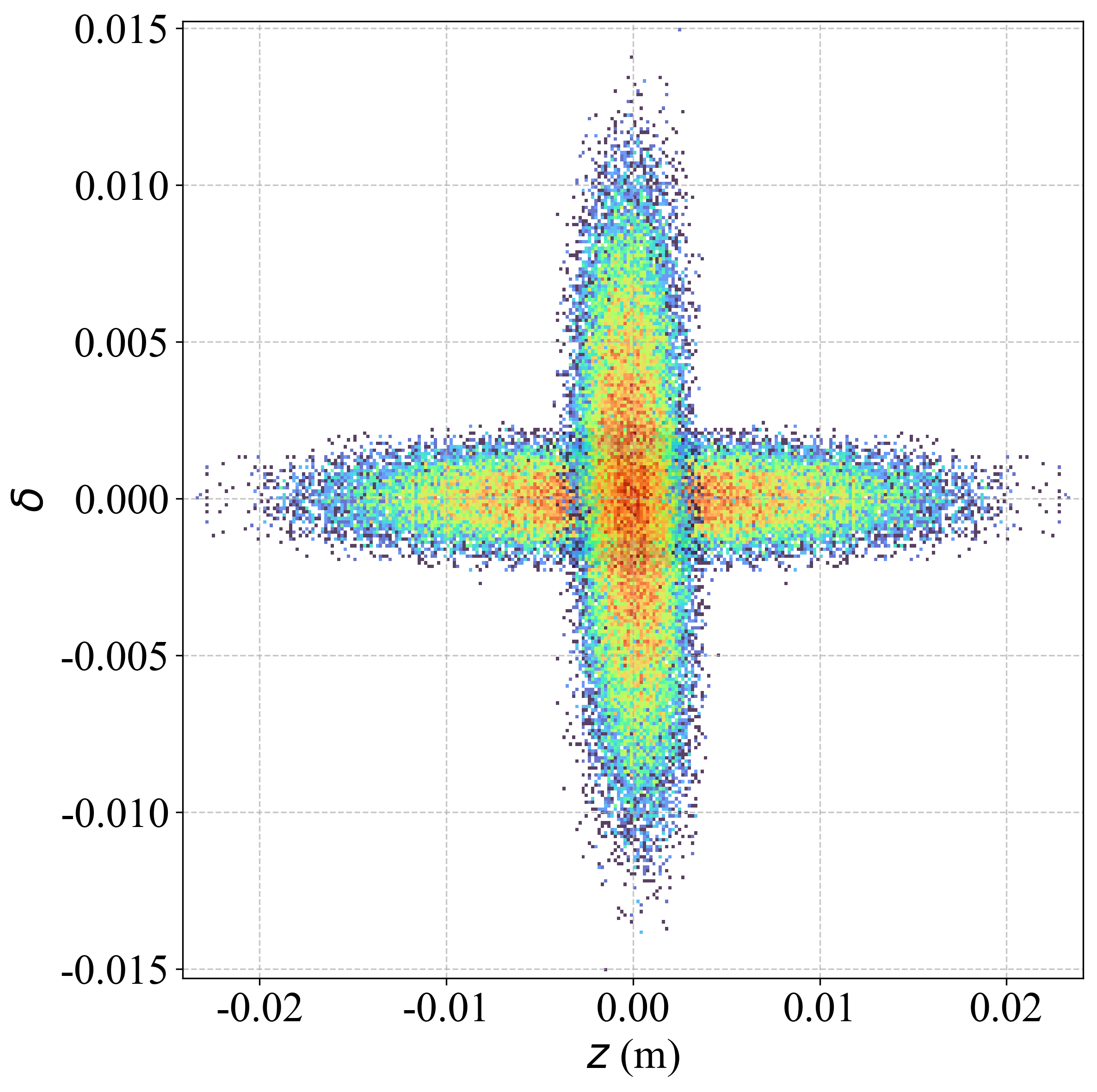} 
\caption{	 Longitudinal phase space before and after bunch compression. }
\label{t_p_density_map}
\end{figure}

\begin{figure}[!h]
\centering
\includegraphics[width=0.42\textwidth]{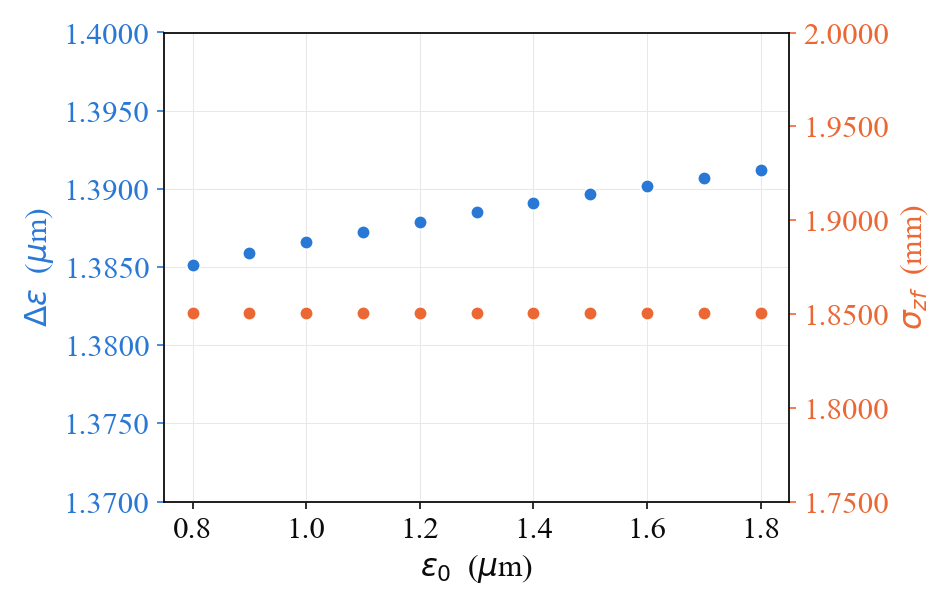}
\includegraphics[width=0.42\textwidth]{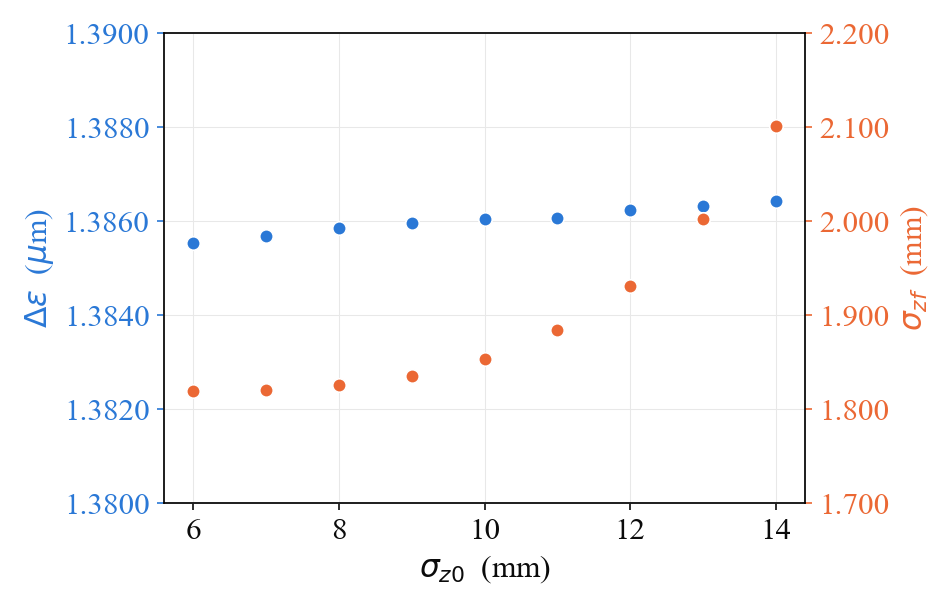}  
\caption{	The  bunch length and projected emittance dilution after bunch compression over a range of initial emittances and bunch lengths.  }
\label{fig2_scan_sigma_s}
\end{figure}

 The detailed chicane parameters for achieving this $R_{56}$ are listed in Table~\ref{tab:chicane_params}. Figure~\ref{t_p_density_map} compares the longitudinal phase space before and after compression.
 Note that the nonlinear distortion of the compressed bunch caused by the   $T_{566}$  term is not significant, because the compression here is attained with a relatively weak RF voltage and   strong magnetic chicanes. Furthermore, the beam extracted from a DLSR exhibits shot-to-shot fluctuations in practice. To verify the robustness of our design against such variations, we scanned the initial normalized emittance and bunch length over a wide range. As shown in Fig. \ref{fig2_scan_sigma_s}, the compressed bunch length remains close to $\sim$2 mm and the emittance growth stays well controlled across the entire scan, demonstrating that our design is robust under realistic operating conditions.

\begin{table}[htbp]
\centering
\caption{Parameters of a single chicane  and the compressed beam parameters.}
\label{tab:chicane_params} 
\begin{tabular*}{10cm} {@{\extracolsep{\fill} } lcc}
\toprule
Parameter & Symbol & Value \\
\midrule
Longitudinal dispersion & $R_{56}$  & $-0.326$\,m \\
Total length & $L_{\mathrm{tot}}$ & 30\,m \\
Dipole length & $L_{\text{B}}$ & 3.0\,m \\
Dipole angle & $\theta $ & $6.98^{\circ}$ \\
Length of 1st (3rd) drift & $L_{\text{d1}}$  ($L_{\text{d3}}$) & 9\,m \\
Length of 2nd drift & $L_{\text{d2}}$ & 0.1\,m \\
\midrule
Final bunch length & $\sigma_{\text{z,f}}$ & 1.85 mm \\
Final normalized emittance & $\varepsilon_{\text{f}}$ & 2.33\,$\mu$m$\cdot$rad \\
Emittance growth ratio & $\Delta\varepsilon / \varepsilon_{0}$ & 148\% \\
Final energy spread & $\sigma_{\delta,\text{f}}$ & 0.59\% \\
\bottomrule
\end{tabular*}
\end{table}

\section{Beam quality and FEL performance}
\label{sec:beam_quality}
 This section evaluates the FEL performance and the beam-quality evolution under the multi-turn radiation scheme. Section~\ref{sec:fw_power} presents a 3D FEL gain analysis based on Xie's  with the TGU modification~\cite{Xie:1995}, quantifying the achievable radiation power and the FEL-induced energy spread growth and vertical emittance dilution. Section~\ref{sec:equilibrium} examines the equilibrium beam parameters reached under the cyclic extraction--compression--radiation--decompression--reinjection--damping process. Section~\ref{sec:photon} determines the achievable photon number per pulse and identifies the optimal compression factor and undulator length. These analytical estimates are validated against 3D time-dependent simulations with \textsc{Genesis} \cite{GENESIS}.

\subsection{Beam quality in the TGU }
\label{sec:fw_power}
The FEL peak power is estimated from a 3D gain analysis. For given beam parameters,
the 1D FEL (Pierce) parameter, the TGU-modified Pierce parameter, and the saturation power are written as~\cite{Xie:1995,DiMitri:2021}
\begin{equation}
\begin{aligned}
\rho_{\mathrm{1D}}
  &= \frac{1}{2\gamma}\left(\frac{I}{I_A}\right)^{1/3}
     \left(\frac{\lambda_u a_w[JJ]}{2\pi\sigma_T}\right)^{2/3},
\\
\rho_{\mathrm{TGU}}
  &= \frac{\rho_{\mathrm{1D}}}{\bigl[1+(\bar\eta_y\sigma_{\delta,u}/\bar\sigma_{y,\beta})^2\bigr]^{1/6}},
\\
P_{s,\mathrm{TGU}}
  &= 1.6\,\rho_{\mathrm{TGU}}P_e\Bigl(\frac{\rho_{\mathrm{TGU}}}{\rho_{\mathrm{1D}}}\Bigr)^2.
\end{aligned}
\end{equation}\label{eq:Psat}
Here   $\sigma_T=\sqrt{\sigma_x^2+\sigma_y^2}$ are  the
total rms transverse size, $I_A\approx 17.05$~kA is the Alfv\'en current, 
$a_w=K/\sqrt{2}$, $[JJ]=J_0(\xi)-J_1(\xi)$ with $\xi=K^2/(4+2K^2)$, 
$\bar\eta_y$ and $\bar\sigma_{y,\beta}$ are the average vertical dispersion and rms 
betatron size along the undulator, $\sigma_{\delta,u}$ is the uncorrelated energy
spread at the undulator entrance, and $P_e=EI/e$ is the electron beam peak power.
Equation~\eqref{eq:Psat} gives the \emph{saturation} power; because the present scheme
operates far from saturation (see below), the single-pass power actually extracted
from the beam is orders of magnitude smaller than $P_{s,\mathrm{TGU}}$.

The single-pass power growth along a TGU of length $L_u$ is described by the logistic
equation for SASE amplification~\cite{DiMitri:2021}
\begin{equation}
P_{\mathrm{FEL}}(n)
  =\frac{P_0\,\exp\bigl(L_u/L_{g,\mathrm{TGU}}\bigr)}
        {1+(P_0/P_{s,\mathrm{TGU}})\exp\bigl(L_u/L_{g,\mathrm{TGU}}\bigr)},
\label{eq:Plogistic}
\end{equation}
where $n$ labels the turn through the TGU, $L_{g,\mathrm{TGU}}$ is the
TGU-modified gain length, and
\begin{equation}
P_0 = \frac{2\pi c e\,\rho_{\mathrm{TGU}}^2 E}{\lambda_s}
\label{eq:P0}
\end{equation}
is the effective shot-noise power at the undulator entrance~\cite{DiMitri:2021}.
In the exponential regime far from saturation, Eq.~\eqref{eq:Plogistic} reduces to
$P_{\mathrm{FEL}}\simeq P_0\,e^{L_u/L_{g,\mathrm{TGU}}}$.

Two principal mechanisms degrade the beam quality during bunch passage
through a TGU: FEL-induced energy spread growth, and vertical emittance
dilution in the dispersive plane.  For the $j$-th TGU ($j=1,\dots,7$) on
the $n$-th turn ($n=1,\dots,24$), the energy spread and vertical emittance  increase  by \cite{Xie:1995}:
\begin{equation}
\begin{aligned}
\Delta\sigma_{\delta,\mathrm{FEL}}^{2}(n,j)
 &=1.6\,\rho_{\mathrm{TGU}}(n,j)\,
  \frac{P_{\mathrm{FEL}}(n,j)}{P_{e}}, \\
\Delta\varepsilon_{y}^{2}(n,j)
& =\varepsilon_{y} (n,j-1)  
  \frac{\bar\eta_{y}^{2}\,\Delta\sigma_{\delta,\mathrm{FEL}}^{2}(n,j )} { \beta_{y}} ,
\end{aligned}
\label{eq:epsyTGU}
\end{equation}
so that
\begin{equation}
\begin{aligned}
\sigma_{\delta}^{2}(n,j)
 &=\sigma_{\delta}^{2}(n,j-1)+\Delta\sigma_{\delta,\mathrm{FEL}}^{2}(n,j), \\
\varepsilon_{y}^{2}(n,j)
 & =\varepsilon_{y}^{2}(n,j-1)  + \Delta\varepsilon_{y}^{2}(n,j)    
\end{aligned}\label{eq:epTGU}
\end{equation}
where $\rho_{\mathrm{TGU}}(n,j-1)$ and $P_{\mathrm{FEL}}(n,j-1)$ are
evaluated with the energy spread and emittance of the beam entering that
TGU.  The entrance state of each turn is the exit of the previous one,
$\sigma_{\delta}(n,0)\equiv\sigma_{\delta}(n-1,7)$ and
$\varepsilon_{y}(n,0)\equiv\varepsilon_{y}(n-1,7)$, with
$(\sigma_{\delta}(1,0),\varepsilon_{y}(1,0))$ the compressed beam.

\subsection{Beam at equilibrium}
\label{sec:equilibrium}

\begin{figure}[!h]
\centering
\includegraphics[width=0.53\textwidth]{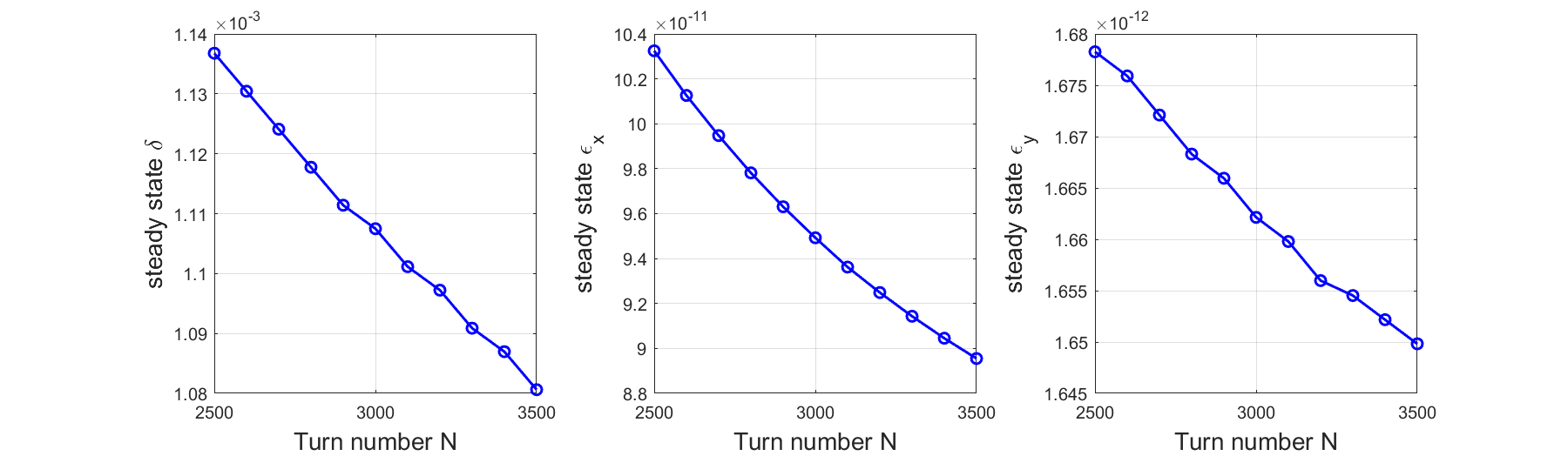} 
\caption{Steady-state beam emittance and energy spread as functions of the damping time in storage ring.   }
\label{damping_time}
\end{figure}

In the cyclic process of extraction, compression, multi-turn radiation,
decompression, reinjection, and damping, the beam quality evolves under three
competing factors: degradation during compression and decompression, turn-by-turn
diffusion in the isochronous ring, and FEL radiation in the TGU. These mechanisms
accumulate over successive turns, whereas radiation damping in the DLSR acts to
restore the beam quality. Their competition drives the beam toward a new steady
state~\cite{DiMitri:2021}, 
\begin{equation}
\begin{aligned}
\varepsilon_x(t) &= \varepsilon_{x,\mathrm{in}}(t)\,e^{-2t/\tau_x} + \varepsilon_{x,\mathrm{eq}}\bigl(1 - e^{-2t/\tau_x}\bigr), \\
\varepsilon_y(t) &= \varepsilon_{y,\mathrm{in}}(t)\,e^{-2t/\tau_y} + \varepsilon_{y,\mathrm{eq}}\bigl(1 - e^{-2t/\tau_y}\bigr),\\
\sigma_\delta^2(t) &= \sigma_{\delta,\mathrm{in}}^2(t)\,e^{-2t/\tau_z} + \sigma_{\delta,\mathrm{eq}}^2\bigl(1 - e^{-2t/\tau_z}\bigr),
\end{aligned}
\end{equation}
where $\varepsilon_{x,\mathrm{in}}$, $\varepsilon_{y,\mathrm{in}}$, $\sigma_{\delta,\mathrm{in}}$ are the beam parameters after accounting for the degradation mechanisms described above; $\varepsilon_{x,\mathrm{eq}}$, $\varepsilon_{y,\mathrm{eq}}$, $\sigma_{\delta,\mathrm{eq}}$ are the equilibrium parameters of the storage ring; and $\tau_x$, $\tau_y$, $\tau_z$ are the radiation damping times in the three degrees of freedom.  
Figure~\ref{damping_time} shows the steady-state beam parameters as functions of the damping time in unit of turns, obtained by solving the evolution equations for a 50~m undulator.

\subsection{Photon number per pulse}
\label{sec:photon}

\begin{figure}
\centering
\includegraphics[width=0.37\textwidth]{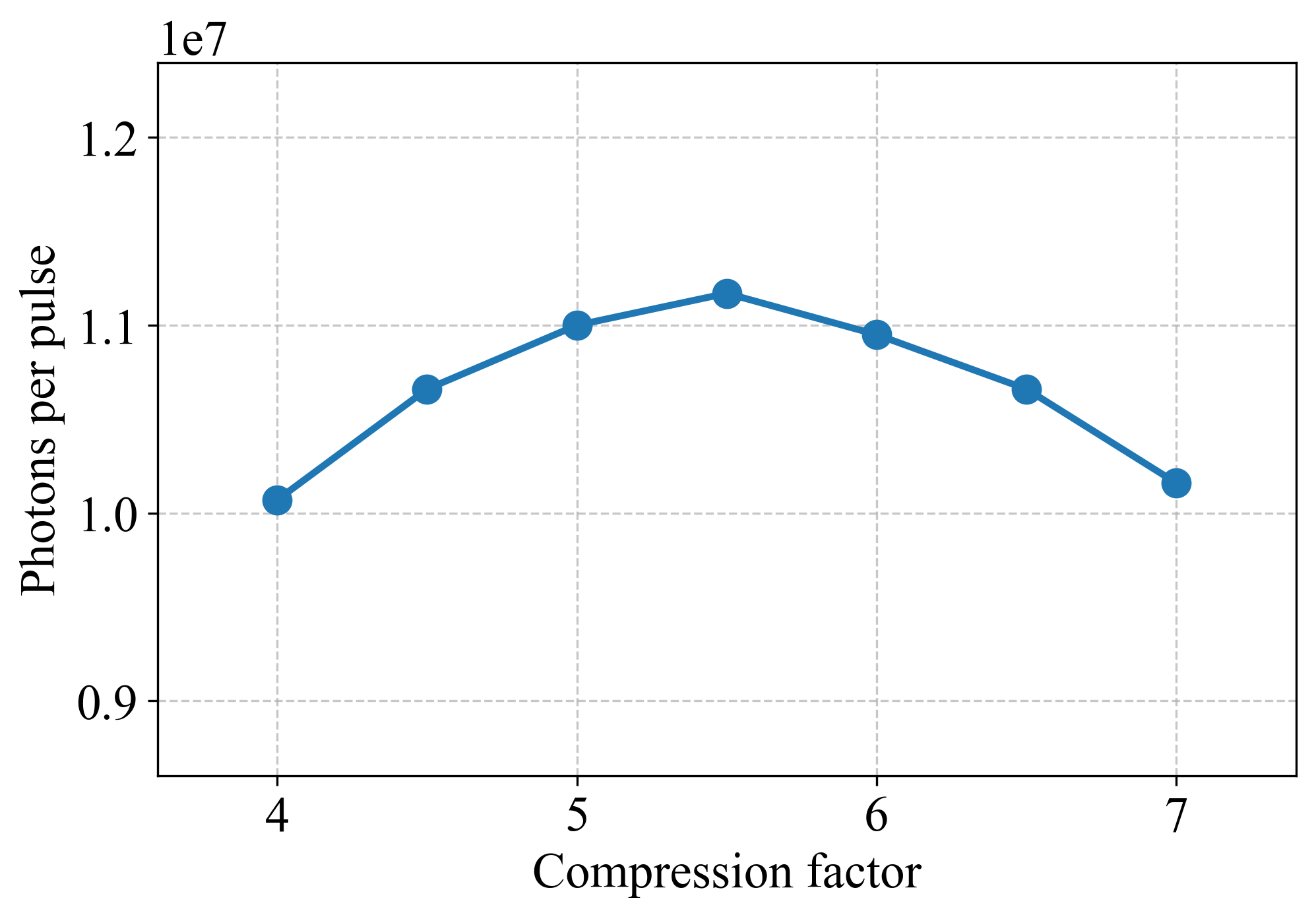} 
\includegraphics[width=0.37\textwidth]{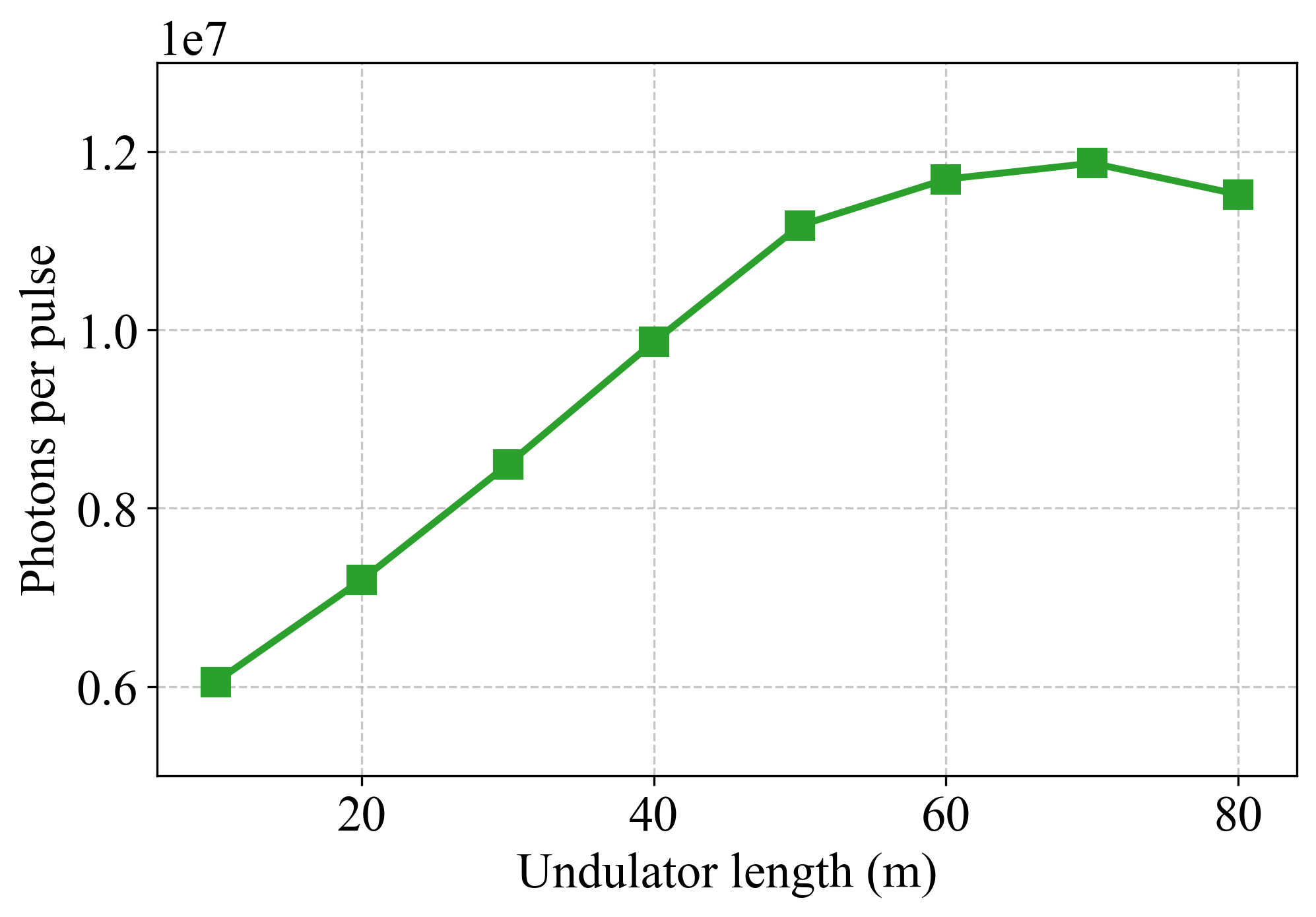} 
\caption{Achievable photon number vs   compression factor $C$  and undulator length.   }
\label{undulator}
\end{figure}

The number of photons per pulse is estimated as
\begin{equation}
N_{\mathrm{ph}} = \frac{P_{\mathrm{FEL}}\,\tau_{\mathrm{pulse}}}{E_{\mathrm{ph}}}  ,
\end{equation}
where $\tau_{\mathrm{pulse}}$ is the pulse duration (approximated by the electron bunch length) and $E_{\mathrm{ph}}$ the single-photon energy.  
Figure \ref{undulator} shows how the achievable photon number changes with the compression factor and undulator length. 
The underlying physics is discussed below.

\begin{figure}[!h]
\centering
\includegraphics[width=0.55\textwidth]{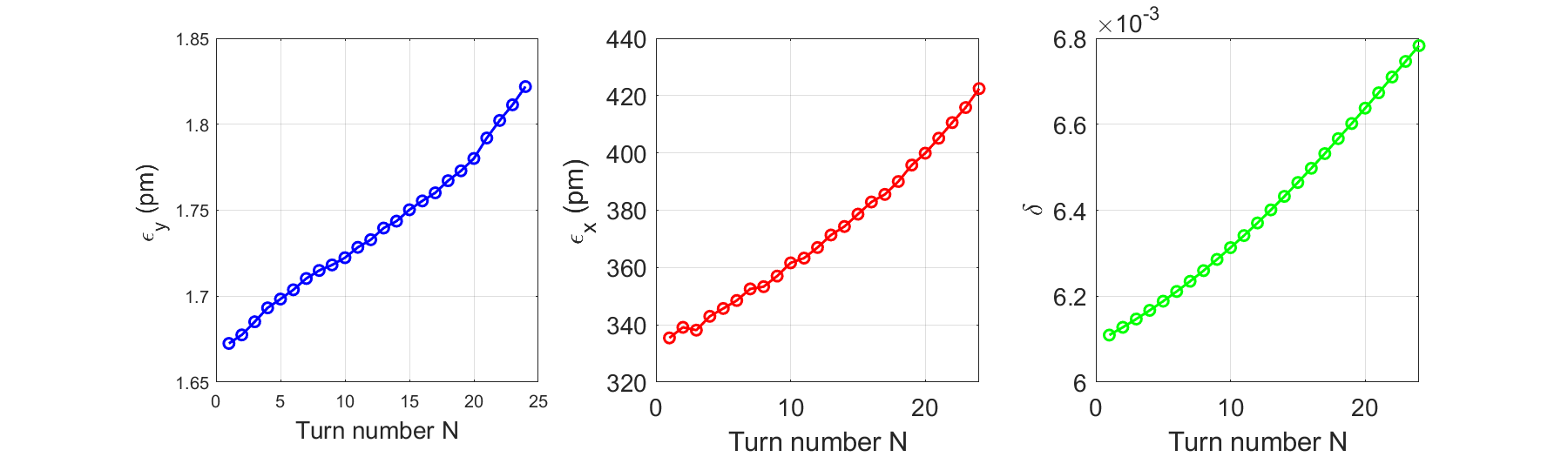}   
\caption{Turn-by-turn evolution of  the beam quality, including horizontal emittance, vertical emittance, and energy spread.  }
\label{beam}
\end{figure}

\begin{figure}[!h]
\centering
\includegraphics[width=0.5\textwidth]{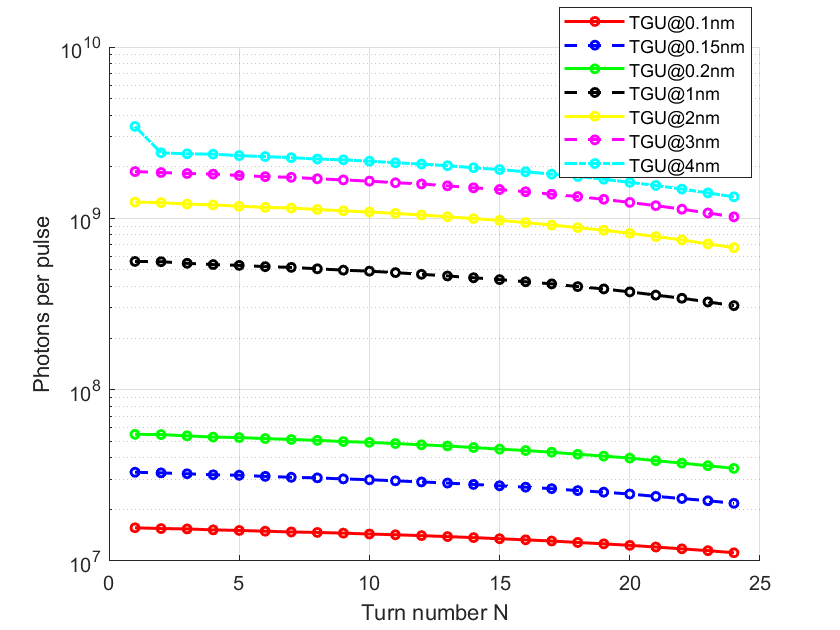}  
\caption{ Turn-by-turn evolution of  the photon number from each TGU.   }
\label{stdeady}
\end{figure}

\begin{figure*}[!h]
\centering
\includegraphics[width=0.9\textwidth]{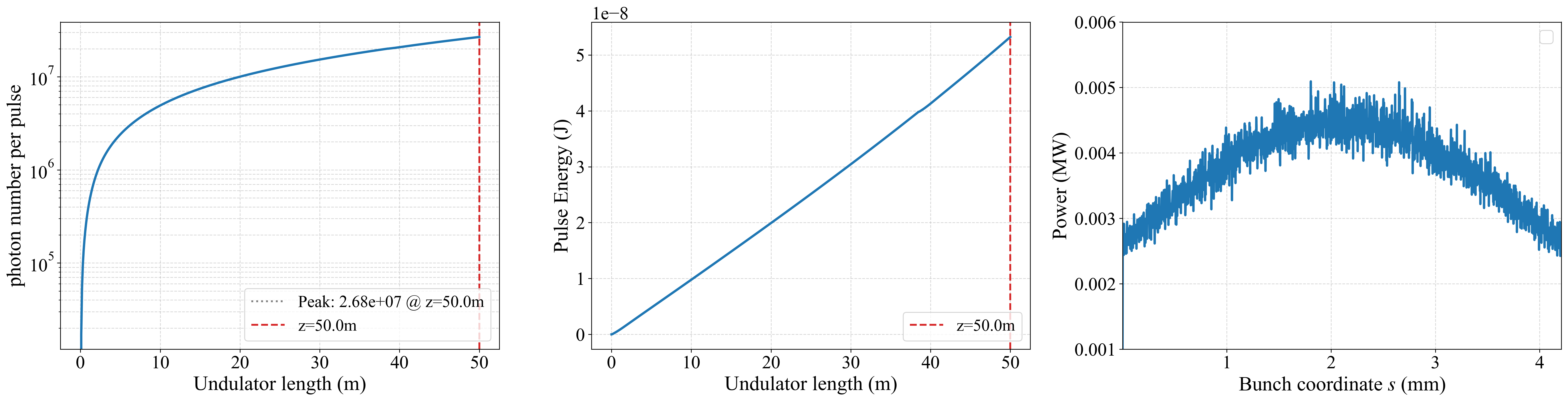}  
\caption{Photon pulse profile and SASE spectrum at a wavelength of 0.1 nm for the 24th-turn FEL. The driving beam has an rms bunch length of 2.1 mm, a peak current of 76 A, a relative energy spread of 0.68\%, and horizontal and vertical emittances of 422 and 1.82 pm.  }
\label{combined_photon_bunching}
\end{figure*} 

At low $C$ ($C<5.5$), the output power is mainly limited by the  peak current. As compression increases, the energy spread grows and the system becomes increasingly sensitive to nonlinear effects, causing the photon number to fall at higher $C$.
 Moreover, the SASE power and photon number increase with the undulator length. Beyond 50~m, however, the
photon output grows   slowly and may even decrease. This is because the radiation-induced
beam degradation accumulates over successive turns and progressively limits the output.

We therefore adopt 50 m undulators and a compression factor of $C = 5.5$. Figure \ref{beam} shows the turn-by-turn evolution of the beam parameters in the isochronous ring. Owing to the accumulated degradation, the photon number decreases turn by turn, as shown in Fig. \ref{stdeady}. Even so, the scheme delivers $N_{\mathrm{ph}} = 1.12 \times 10^{7}$--$1.56 \times 10^{7}$ photons per pulse at 0.1 nm over the multi-turn radiation, and $N_{\mathrm{ph}} >  1.34  \times 10^{9}$ at 4~nm.

 To validate the analytical results described above, we extract the beam
parameters at each turn and perform three-dimensional time-dependent
simulations with \textsc{Genesis}~\cite{GENESIS}. For a 50 m undulator at
0.1 nm, the simulation yields \(2.68\times10^{7}\) photons per pulse, i.e., a
pulse energy of 0.05~\(\mu\)J, with a gain length of 11.95 m. The 50 m
undulator spans about 4.2 gain lengths, so the FEL remains far from
saturation. This agrees with the analytical prediction within an order of magnitude and is somewhat higher.   The photon pulse profile and the SASE spectrum
are shown in Fig.~\ref{combined_photon_bunching}.

\section{Discussion and Conclusions}
\label{sec:summary}
The dual-ring light source expands the capabilities of DLSRs into the FEL regime while preserving the advantages of a storage-ring. The scheme achieves MHz-level FEL repetition rates through  the multi-turn radiation strategy in the isochronous ring. With a pulse duration of a few ps, the dual-ring source delivers $ >  10^{7}$ photons per pulse at \text{0.1~nm} and $>  10^{9}$ photons per pulse at 4 nm. This combination of ps temporal resolution and high photon flux bridges a critical gap between DLSRs and linac-driven X-ray FELs, meeting the requirements of time-resolved fine spectroscopic analysis and related ps-scale applications.

The dual-ring also provides considerable operational flexibility. At the FEL
high-repetition-rate limit of  MHz-level, the scheme operates as described above. The repetition
rate is tunable from 0.1~kHz to 1~MHz by adjusting the kicker repetition rate and the number of turns in the isochronous
ring. At lower  FEL repetition rates, for instance in single-pass radiation, it delivers
$\sim$1~ps pulses with $\sim$100~MW peak power, comparable to the bypass scheme of
Di~Mitri \textit{et al.}~\cite{DiMitri:2021}.

The scheme is also compatible with modifications of the undulator line for improved longitudinal coherence. An externally seeded configuration, in which the FEL starts from a coherent seed laser rather than from shot noise, would yield superior longitudinal coherence over the standard SASE mode~\cite{Feng:2017,Yang:2024}. The XFELO configuration, which augments the SASE undulator with X-ray optical cavities to accumulate gain over multiple round trips, is also viable~\cite{HuaYu}. The high-repetition-rate bunch delivery in the isochronous ring makes narrow-bandwidth operation at short wavelengths a realistic prospect.

Unlike a conventional storage ring, the isochronous ring has no RF system of its own; nevertheless, the energy lost by the beam to synchrotron radiation
can be compensated downstream. 
The estimated loss in the TBA dipoles is about
80 MeV per electron (1.33\% of 6 GeV) over the 24 lasing turns.  This
loss can be compensated by the decompression system, so that the beam sees an
accelerating phase instead of the zero crossing.

 The scheme also imposes stringent requirements on the timing and synchronization system. The extraction kicker must be locked to the DLSR bunch pattern so that exactly five adjacent bunches are removed from their buckets; the compressor and
decompressor RF must be phase-locked to the arrival of the batch; and the batch must be reinjected into the same buckets from which it was extracted.
This last requirement implies that the total path length---the 24 turns in the
isochronous ring plus the two transfer lines---must be an integer multiple of
the DLSR circumference. A simple estimate gives \(C_{\rm tr}=284\) m for each
of the two transfer lines, so that \(2C_{\rm tr}+24C_{\rm iso}=20C_{\rm ring}\).

 The injection and extraction system based on stripline kickers still presents a challenge to current state-of-the-art technology.
  For a 6 GeV beam with a 26.7 ns pulse width operating at 10 kHz,  together with rise and fall times  about 6~ns,  
the estimated average kicker power reaches the tens of kilowatts, beyond the level demonstrated in existing facilities. Further engineering studies are required to validate this subsystem.

This work presents the conceptual design of the dual-ring light source and establishes its framework. The present study is idealized in several respects: beam losses during transport have not been studied in detail, collective effects in the isochronous ring, 
such as microbunching instability and wakefields,  have not been considered, and a complete start-to-end simulation has not yet been performed. Such a simulation would require a unified code capable of modeling both FEL and storage-ring beam dynamics. Future work will incorporate more realistic physical effects, advancing toward experimental demonstration.

The definitive layout is beyond the scope of the present study and will be
dictated by the site constraints. At HEPS, for instance, the space enclosed by
the storage ring is occupied by a injector linac and a booster,
so that a concentric arrangement of the isochronous ring is not practicable; a
parallel configuration is accordingly preferred, which also permits construction
and commissioning of the new ring without interrupting user operation of the
DLSR.

\bibliography{iucr}

@article{Tavares:2018,
  author = "P. F. Tavares and S. C. Leemann and M. Sj{\"o}str{\"o}m and A. Andersson",
  title = "Commissioning and first operation of the MAX-IV 3\,GeV storage ring",
  journal = "J. Synchrotron Rad.",
  volume = 25,
  pages = "1291--1302",
  doi = "10.1107/S1600577518008111",
  year = 2018,
}

@article{Sajaev:2019,
  author = "V. Sajaev",
  title = "Commissioning simulations for the Argonne Advanced Photon Source upgrade lattice",
  journal = "Phys. Rev. Accel. Beams",
  volume = 22,
  pages = "040102",
  doi = "10.1103/PhysRevAccelBeams.22.040102",
  year = 2019,
}

@article{Jiao:2018,
  author = "Y. Jiao and G. Xu and X. H. Cui and others",
  title = "The HEPS project",
  journal = "J. Synchrotron Rad.",
  volume = 25,
  pages = "1611--1618",
  doi = "10.1107/S1600577518012110",
  year = 2018,
}

@article{Bai:2026,
  author = "Z. Bai and B. Wei and X. Liu and others",
  title = "Hefei Advanced Light Facility: a soft X-ray diffraction-limited storage ring light source in China",
  journal = "Photon Sci.",
  doi = "10.1021/photonsci.6c00037",
  year = 2026,
}

@article{Raimondi:2023,
  author = "P. Raimondi and C. Benabderrahmane and P. Berkvens and others",
  title = "The Extremely Brilliant Source storage ring of the European Synchrotron Radiation Facility",
  journal = "Commun. Phys.",
  volume = 6,
  pages = "82",
  doi = "10.1038/s42005-023-01195-z",
  year = 2023,
}

@article{Liu:2014,
  author = "L. Liu and N. Milas and A. H. C. Mukai and X. R. Resende and F. H. de S{\'a}",
  title = "The Sirius project",
  journal = "J. Synchrotron Rad.",
  volume = 21,
  pages = "904--911",
  doi = "10.1107/S1600577514011928",
  year = 2014,
}

@article{Emma:2010,
  author = "P. Emma and R. Akre and J. Arthur and others",
  title = {First lasing and operation of an {\aa}ngstrom-wavelength free-electron laser},
  journal = "Nat. Photon.",
  volume = 4,
  pages = "641--647",
  doi = "10.1038/nphoton.2010.176",
  year = 2010,
}

@article{Ishikawa:2012,
  author = "T. Ishikawa and H. Aoyagi and T. Asaka and others",
  title = {A compact X-ray free-electron laser emitting in the sub-{\aa}ngstr{\"o}m region},
  journal = "Nat. Photon.",
  volume = 6,
  pages = "540--544",
  doi = "10.1038/nphoton.2012.141",
  year = 2012,
}

@article{Hara:2013,
  author = "T. Hara",
  title = "Fully coherent soft X-rays at FERMI",
  journal = "Nat. Photon.",
  volume = 7,
  pages = "852--854",
  doi = "10.1038/nphoton.2013.279",
  year = 2013,
}

@article{Allaria:2012,
  author = "E. Allaria and R. Appio and L. Badano and others",
  title = "Highly coherent and stable pulses from the FERMI seeded free-electron laser in the extreme ultraviolet",
  journal = "Nat. Photon.",
  volume = 6,
  pages = "699--704",
  doi = "10.1038/nphoton.2012.233",
  year = 2012,
}

@article{Kang:2017,
  author = "H. S. Kang and C. K. Min and H. Heo and others",
  title = "Hard X-ray free-electron laser with femtosecond-scale timing jitter",
  journal = "Nat. Photon.",
  volume = 11,
  pages = "708--713",
  doi = "10.1038/s41566-017-0029-8",
  year = 2017,
}

@article{Prat:2020,
  author = "E. Prat and R. Abela and M. Aiba and others",
  title = "A compact and cost-effective hard X-ray free-electron laser driven by a high-brightness and low-energy electron beam",
  journal = "Nat. Photon.",
  volume = 14,
  pages = "748--754",
  doi = "10.1038/s41566-020-00712-8",
  year = 2020,
}

@inproceedings{SHINE:2018,
  author = "Z. Zhao and D. Wang and Z. Yang and L. Yin",
  title = "SCLF: an 8-GeV CW SCRF linac-based X-ray FEL facility in Shanghai",
  booktitle = "Proc. 38th International Free Electron Laser Conference (FEL'17)",
  address = "Santa Fe, NM, USA",
  publisher = "JACoW",
  doi = "10.18429/JACoW-FEL2017-MOP055",
  year = 2018,
}

@inproceedings{Galayda:2018,
  author = "J. Galayda",
  title = "The LCLS-II: a high power upgrade to the LCLS",
  booktitle = "Proc. 9th International Particle Accelerator Conference (IPAC'18)",
  address = "Vancouver, Canada",
  publisher = "JACoW",
  note = "Paper MOYGB2",
  doi = "10.18429/JACoW-IPAC2018-MOYGB2",
  year = 2018,
}

@inproceedings{Raubenheimer:2018,
  author = "T. Raubenheimer",
  title = "The LCLS-II-HE, a high energy upgrade of the LCLS-II",
  booktitle = "Proc. 60th ICFA Advanced Beam Dynamics Workshop on Future Light Sources (FLS2018)",
  address = "Shanghai, China",
  publisher = "JACoW",
  doi = "10.18429/JACoW-FLS2018-MOP1WA02",
  year = 2018,
}

@article{Liu:2023,
  author = "S. Liu and C. Grech and M. Guetg and others",
  title = "Cascaded hard X-ray self-seeded free-electron laser at megahertz repetition rate",
  journal = "Nat. Photon.",
  volume = 17,
  pages = "984--991",
  doi = "10.1038/s41566-023-01305-x",
  year = 2023,
}

@article{Allaria:2013,
  author = "E. Allaria and D. Castronovo and P. Cinquegrana and others",
  title = "Two-stage seeded soft-X-ray free-electron laser",
  journal = "Nat. Photon.",
  volume = 7,
  pages = "913--918",
  doi = "10.1038/nphoton.2013.277",
  year = 2013,
}

@article{Roseker:2020,
  author = "W. Roseker and S. O. Hruszkewycz and F. Lehmk{\"u}hler and others",
  title = "Towards ultrafast dynamics with split-pulse X-ray photon correlation spectroscopy at free electron laser sources",
  journal = "Nat. Commun.",
  volume = 9,
  pages = "1704",
  doi = "10.1038/s41467-018-04178-9",
  year = 2018,
}

@article{Chapman:2011,
  author = "H. N. Chapman and P. Fromme and A. Barty and others",
  title = "Femtosecond X-ray protein nanocrystallography",
  journal = "Nature",
  volume = 470,
  pages = "73--77",
  doi = "10.1038/nature09750",
  year = 2011,
}

@article{Seibert:2011,
  author = "M. M. Seibert and T. Ekeberg and F. R. N. C. Maia and others",
  title = "Single mimivirus particles intercepted and imaged with an X-ray laser",
  journal = "Nature",
  volume = 470,
  pages = "78--81",
  doi = "10.1038/nature09748",
  year = 2011,
}

@article{Fratalocchi:2011,
  author = "A. Fratalocchi and G. Ruocco",
  title = "Single-molecule imaging with X-ray free-electron lasers: dream or reality?",
  journal = "Phys. Rev. Lett.",
  volume = 106,
  pages = "105504",
  doi = "10.1103/PhysRevLett.106.105504",
  year = 2011,
}

@article{Li:2024,
  author = "Z. Li and Y. Fu and others",
  title = "Roaming in highly excited states: the central atom elimination of triatomic molecule decomposition",
  journal = "Science",
  volume = 383,
  pages = "746--750",
  doi = "10.1126/science.adn3357",
  year = 2024,
}

@article{Huang:2025,
  author = "E. Huang and L. Zhong and others",
  title = {Conformer-specific infrared spectroscopy of cationic Criegee intermediates syn- and anti-CH$_3$CHOO$^+$},
  journal = "Nat. Commun.",
  volume = 16,
  pages = "2313",
  doi = "10.1038/s41467-025-57670-4",
  year = 2025,
}

@article{Billardon:1983,
  author = "M. Billardon and P. Elleaume and J. M. Ortega and C. Bazin and M. Bergher and M. Velghe and D. A. G. Deacon and Y. Petroff",
  title = "First operation of a storage-ring free-electron laser",
  journal = "Phys. Rev. Lett.",
  volume = 51,
  pages = "1652--1655",
  doi = "10.1103/PhysRevLett.51.1652",
  year = 1983,
}

@article{Couprie:1991,
  author = "M. E. Couprie and M. Velghe and R. Prazeres and D. Jaroszynski and J. M. Ortega",
  title = "Operation of the Super-ACO storage ring free electron laser in the UV",
  journal = "Nucl. Instrum. Methods Phys. Res., Sect. A",
  volume = 304,
  pages = "47--52",
  doi = "10.1016/0168-9002(91)90820-G",
  year = 1991,
}

@article{XFELO,
  author = "Y. Li and R. Lindberg and K.-J. Kim",
  title = "Transverse gradient undulator in a storage ring X-ray free electron laser oscillator",
  journal = "Phys. Rev. Accel. Beams",
  volume = 26,
  pages = "030702",
  doi = "10.1103/PhysRevAccelBeams.26.030702",
  year = 2023,
}

@article{HuaYu,
  author = "L. H. Yu and V. Smaluk and T. Shaftan and G. Tiwari and X. Yang",
  title = "3D small-gain formula allowing strong focusing and harmonic lasing for a ring-based X-ray free electron laser oscillator",
  journal = "Phys. Rev. Accel. Beams",
  volume = 27,
  pages = "060702",
  doi = "10.1103/PhysRevAccelBeams.27.060702",
  year = 2024,
}

@article{ADM,
  author = "C. Li and B. Jiang and C. Feng and Q. Gu and Z. Bai and W. Wan and Q. Zhang and K. Wang",
  title = "Toward a storage ring coherent light source based on an angular dispersion-induced microbunching scheme",
  journal = "J. Synchrotron Rad.",
  volume = 32,
  pages = "82--89",
  doi = "10.1107/S1600577524010907",
  year = 2025,
}

@article{SSMB,
  author = "X. J. Deng and Y. Zhang and Z. L. Pan and Z. Z. Li and J. H. Bian and C.-Y. Tsai and R. K. Li and A. W. Chao and W. H. Huang and C. X. Tang",
  title = "Average and statistical properties of coherent radiation from steady-state microbunching",
  journal = "J. Synchrotron Rad.",
  volume = 30,
  pages = "35--50",
  doi = "10.1107/S1600577522009973",
  year = 2023,
}

@article{xiaofancai,
  author = "X. Wang and C. Feng and T. Liu and Z. Zhang and C.-Y. Tsai and J. Wu and C. Yang and Z. Zhao",
  title = "Angular dispersion enhanced prebunch for seeding ultrashort and coherent EUV and soft X-ray free-electron laser in storage rings",
  journal = "J. Synchrotron Rad.",
  volume = 26,
  pages = "677--684",
  doi = "10.1107/S1600577519002674",
  year = 2019,
}

@article{Liu:2026,
  author = "W. Liu and Y. Zhao and W. Qin and Y. Jiao and X. Li and S. Wang",
  title = "High-harmonic coherent pulse generation in a storage ring using multiple-echo-enabled harmonic generation",
  journal = "Phys. Rev. Accel. Beams",
  volume = 29,
  pages = "093405",
  doi = "10.1103/6gc2-41zw",
  year = 2026,
}

@article{DiMitri:2021,
  author = "S. Di Mitri and M. Cornacchia and B. Diviacco and G. Perosa and F. Sottocorona and S. Spampinati",
  title = "Bridging the gap of storage ring light sources and linac-driven free-electron lasers",
  journal = "Phys. Rev. Accel. Beams",
  volume = 24,
  pages = "060702",
  doi = "10.1103/PhysRevAccelBeams.24.060702",
  year = 2021,
}

@article{Zhou:2023,
  author = "K. Zhou and R. Li and Z. Bai",
  title = "Sustainable early-stage lasing in a low-emittance electron storage ring",
  journal = "Appl. Phys. Lett.",
  volume = 123,
  pages = "241106",
  doi = "10.1063/5.0177748",
  year = 2023,
}

@article{Cai:2025,
  author = "Y. Cai and J. Wu",
  title = "Beam dynamics in a storage ring with a free-electron laser",
  journal = "Phys. Rev. Accel. Beams",
  volume = 28,
  pages = "050701",
  doi = "10.1103/PhysRevAccelBeams.28.050701",
  year = 2025,
}

@article{Feng:2017,
  author = "C. Feng and Z. Zhao",
  title = "A storage ring-based free-electron laser for generating ultrashort coherent EUV and X-ray radiation",
  journal = "Sci. Rep.",
  volume = 7,
  pages = "4724",
  doi = "10.1038/s41598-017-04962-5",
  year = 2017,
}

@article{Yang:2024,
  author = "X. Yang and L. Yu and V. Smaluk and T. Shaftan",
  title = "High harmonic generation seeding echo-enabled harmonic generation toward a storage ring-based tender and hard X-ray free-electron laser",
  journal = "Instruments",
  volume = 8,
  pages = "35",
  doi = "10.3390/instruments8020035",
  year = 2024,
}

@article{RDTM,
  author = "H. Xu and C. Meng and Y. Peng and others",
  title = "Equilibrium electron beam parameters of the High Energy Photon Source",
  journal = "Radiat. Detect. Technol. Methods",
  volume = 7,
  pages = "279--287",
  doi = "10.1007/s41605-022-00374-w",
  year = 2023,
}

@inproceedings{MOPSO,
  author = "C. A. Coello Coello and M. S. Lechuga",
  title = "MOPSO: A proposal for multiple objective particle swarm optimization",
  booktitle = "Proceedings of the 2002 Congress on Evolutionary Computation (CEC'02)",
  address = "Honolulu, HI, USA",
  publisher = "IEEE",
  pages = "1051--1056",
  doi = "10.1109/CEC.2002.1004388",
  year = 2002,
}

@article{elegant,
  author = "M. Borland",
  title = "Simple method for particle tracking with coherent synchrotron radiation",
  journal = "Phys. Rev. ST Accel. Beams",
  volume = 4,
  pages = "070701",
  doi = "10.1103/PhysRevSTAB.4.070701",
  year = 2001,
}

@article{Huang:2012,
  author = "Z. Huang and Y. Ding and C. B. Schroeder",
  title = "Compact X-ray Free-Electron Laser from a Laser-Plasma Accelerator Using a Transverse-Gradient Undulator",
  journal = "Phys. Rev. Lett.",
  volume = 109,
  pages = "204801",
  doi = "10.1103/PhysRevLett.109.204801",
  year = 2012,
}

@article{Halbach:1983,
  author = "K. Halbach",
  title = "Permanent magnet undulators",
  journal = "J. Phys. (Paris), Colloq.",
  volume = 44,
  pages = "C1-211--C1-216",
  year = 1983,
}

@inproceedings{Xie:1995,
  author = "M. Xie",
  title = "Design optimization for an X-ray free electron laser driven by SLAC LINAC",
  booktitle = "Proc. 1995 Particle Accelerator Conference (PAC'95)",
  address = "Dallas, TX, USA",
  pages = "183--185",
  doi = "10.1109/PAC.1995.504603",
  year = 1995,
}

@article{Strickland:1985,
  author = "D. Strickland and G. Mourou",
  title = "Compression of amplified chirped optical pulses",
  journal = "Opt. Commun.",
  volume = 55,
  pages = "447--449",
  doi = "10.1016/0030-4018(85)90151-8",
  year = 1985,
}

@article{Bartosik:2022,
  author = "H. Bartosik and Y. Papaphilippou and A. Wolski",
  title = "A first taste of nonlinear beam dynamics",
  journal = "Proc. CERN Accelerator School: Introduction to Accelerator Physics",
  eprint = "arXiv:2201.01532",
  year = 2022,
}

@article{Hajima,
  author = "R. Hajima",
  title = "Emittance compensation in a return arc of an energy-recovery linac",
  journal = "Nucl. Instrum. Methods Phys. Res., Sect. A",
  volume = 528,
  pages = "335--339",
  doi = "10.1016/j.nima.2004.04.063",
  year = 2004,
}

@article{Zeng:2025,
  author = "F. Zeng and Y. Jiao and W. Liu and C.-Y. Tsai",
  title = "Solvable model of the coherent synchrotron radiation effects in a four-bend chicane compressor",
  journal = "New J. Phys.",
  volume = 27,
  pages = "063801",
  doi = "10.1088/1367-2630/addde2",
  year = 2025,
}

@article{Zeng:20252,
  author = "F. Zeng and Y. Jiao and W. Liu and C.-Y. Tsai",
  title = "Cancelation of coherent synchrotron radiation kicks in chicane bunch compressors",
  journal = "Phys. Rev. Accel. Beams",
  volume = 28,
  pages = "070701",
  doi = "10.1103/2vjn-t4qb",
  year = 2025,
}

@book{Chao:2006,
  author = "A. W. Chao and M. Tigner",
  title = "Handbook of Accelerator Physics and Engineering",
  edition = "3rd",
  publisher = "World Scientific",
  address = "Singapore",
  year = 2006,
}

@article{EEHG2,
  author = "D. Xiang and G. Stupakov",
  title = "Echo-enabled harmonic generation free electron laser",
  journal = "Phys. Rev. ST Accel. Beams",
  volume = 12,
  pages = "030702",
  doi = "10.1103/PhysRevSTAB.12.030702",
  year = 2009,
}

@article{GENESIS,
  author = "S. Reiche",
  title = "GENESIS 1.3: A fully 3D time-dependent FEL simulation code",
  journal = "Nucl. Instrum. Methods Phys. Res., Sect. A",
  volume = 429,
  pages = "243",
  doi = "10.1016/S0168-9002(99)00114-X",
  year = 1999,
}

\end{document}